\documentclass[10pt]{article}
\usepackage{amsfonts}
\usepackage{amsmath}
\usepackage{amssymb} 
\usepackage{amsthm}
\usepackage{braket}                 
\usepackage{commath}                
\usepackage{physics}                
\usepackage[compress]{cite}
\usepackage{bm} 
\usepackage{color}
\usepackage{comment}
\includecomment{comment}
\usepackage{dcolumn}
\usepackage{enumerate}
\usepackage{caption}

\DeclareMathOperator\Erf{Erf}
\DeclareMathOperator\OwenT{OwnT} 
\DeclareMathOperator\Sign{Sign} 

\usepackage{graphicx}
\usepackage{epstopdf}   
\usepackage{mathrsfs}
\usepackage{mathtools} 
\usepackage{paralist}
\usepackage[parfill]{parskip}    
\usepackage{textcomp}
\usepackage{titlesec}
\usepackage{varioref}
\usepackage{hyperref} 
\usepackage[normalem]{ulem}
\allowdisplaybreaks
\hypersetup{
    pdfstartview={FitH},    
    colorlinks=true,       
    linkcolor=blue,          
    citecolor=blue,        
    filecolor=blue,      
    urlcolor=magenta           
}
\DeclareFontFamily{OT1}{pzc}{}
\DeclareFontShape{OT1}{pzc}{m}{it}%
             {<-> s * [0.900] pzcmi7t}{}
\DeclareMathAlphabet{\mathscr}{OT1}{pzc}%
                                 {m}{it}
\labelformat{section}{Section #1} 
\labelformat{subsection}{Section #1} 
\labelformat{equation}{Eq.~(#1)} 
\labelformat{figure}{Fig.~#1} 
\labelformat{subfigure}{Fig.~\thefigure#1} 
\labelformat{table}{Tab.~#1} 
\titleclass{\subsubsubsection}{straight}[\subsection]

\newcounter{subsubsubsection}[subsubsection]
\renewcommand\thesubsubsubsection{\thesubsubsection.\arabic{subsubsubsection}}

\titleformat{\subsubsubsection}
{\normalfont\normalsize\bfseries}{\thesubsubsubsection}{1em}{}
\titlespacing*{\subsubsubsection}
{0pt}{3.25ex plus 1ex minus .2ex}{1.5ex plus .2ex}
\makeatletter
\renewcommand\paragraph{\@startsection{paragraph}{5}{\z@}%
	{3.25ex \@plus1ex \@minus.2ex}%
	{-1em}%
	{\normalfont\normalsize\bfseries}}
\renewcommand\subparagraph{\@startsection{subparagraph}{6}{\parindent}%
	{3.25ex \@plus1ex \@minus .2ex}%
	{-1em}%
	{\normalfont\normalsize\bfseries}}
\def\toclevel@subsubsubsection{4}
\def\toclevel@paragraph{5}
\def\toclevel@paragraph{6}
\def\l@subsubsubsection{\@dottedtocline{4}{7em}{4em}}
\def\l@paragraph{\@dottedtocline{5}{10em}{5em}}
\def\l@subparagraph{\@dottedtocline{6}{14em}{6em}}
\makeatother
\newcommand{\be}{\begin{equation}}
\newcommand{\ee}{\end{equation}}
\newcommand{\bea}{\begin{eqnarray}}
\newcommand{\eea}{\end{eqnarray}}

\def\P[#1]#2{\Pi^{#1}_{\phantom{a}#2}}
\def\Pt[#1]#2{\tilde{\Pi}_{#1}^{\phantom{a}#2}}

\usepackage{appendix}
\usepackage{hyperref}
\usepackage[margin=0.4in]{geometry}
\usepackage{comment}
\usepackage{verbatim}

\hypersetup{
    colorlinks=true,
    linkcolor=blue,
    filecolor=magenta,
    urlcolor=red,
    citecolor=green,
}
\title{Violation of Bell inequality from a squeezed coherent state of inflationary perturbations}
\author{Aurindam Mondal \footnote{aurindammondal99@gmail.com}$~^{1}$ 
,  Rathul Nath Raveendran \footnote{rathulnath.r@gmail.com}$~^{2,3}$ \\
{\small{$^{1}$ Indian Statistical Institute, Kolkata- 700108, India}}\\
{$^{2}$\small{Indian Association for the Cultivation of Science, Kolkata- 700032, India}}\\
{$^{3}$\small{Department of Physics, Indian Institute of Science,
C. V. Raman Road, Bangalore 560012, India}}}

\date{\vspace{-0.6cm} }  

\begin{document}

\maketitle
\abstract{We investigate the quantum nature of primordial perturbations by studying the violation of Bell inequality when the initial state is taken to be a coherent state rather than the usual Bunch-Davies vacuum. As inflation progresses, the coherent state evolves into a squeezed coherent state, and we derive an analytical expression for the expectation value of the Bell operator constructed from pseudo-spin operators. Our analysis shows that although the expectation value of the Bell operator initially deviates from the vacuum case, it asymptotically saturates to the same value. Notably, this saturation occurs more rapidly for non-zero coherent state parameters, indicating that a larger one-point correlation function accelerates the approach to maximal Bell inequality violation.}
\tableofcontents
\section{Introduction}\label{sec:1}

One of the most fascinating and challenging aspects of studying the early universe lies at the intersection of gravity and quantum mechanics—two foundational pillars of physics that remain difficult to reconcile. The extreme conditions of the early universe offer a unique opportunity to test and potentially unify these frameworks. Central questions in this context include understanding the origin of the universe and explaining the emergence of large-scale cosmic structure.

According to the inflationary paradigm, the seeds of large-scale structure we see today originated as quantum vacuum fluctuations during a brief period of rapid, accelerated expansion in the early universe. This expansion stretched the fluctuations beyond the Hubble horizon, effectively freezing them into classical density perturbations. As the universe continued to evolve and these modes re-entered the horizon, they acted as the initial conditions for gravitational collapse in over dense regions—eventually giving rise to galaxies, clusters, and the vast cosmic web we observe today~\cite{Guth:1980zm,Starobinsky:1982ee,Kodama:1984ziu,MUKHANOV1992203}.

Although this mechanism successfully explains key cosmological observations—such as the temperature and polarization anisotropies in the Cosmic Microwave Background (CMB)—its conceptual foundations remain unresolved. In particular, the quantum origin of primordial perturbations is not yet firmly established, as it involves quantizing both metric and matter field fluctuations in the absence of a complete theory of quantum gravity. Moreover, it has been argued that a deeper theoretical and observational understanding of the quantum nature of these primordial perturbations could offer valuable insights into the very foundations of quantum mechanics itself. For these reasons, probing the possible quantum origin of cosmological perturbations remains an important and compelling endeavor ~\cite{Polarski:1995jg,Albrecht:1992kf,Kiefer:1998jk,Martin:2015qta,Martin:2004um,Martin:2012pea,Kanno:2017dci,Raveendran:2022dtb,Raveendran:2023dst}.

In inflationary cosmology, quantum field fluctuations are decomposed into momentum modes whose dynamics are governed by time-dependent mode functions. More precisely, the quantum field can be represented as an infinite collection of independent bipartite systems $\{\mathbf{k}, -\mathbf{k} \}$, where inflationary expansion creates entangled particle pairs with opposite momenta. The standard procedure to extract observable predictions of inflation involves imposing suitable initial conditions on these modes when the modes are deep inside the Hubble horizon, evolving them forward, and then computing the relevant quantum correlation functions of perturbations at the end of inflation. These correlation functions then undergo further evolution during the post-inflationary era, ultimately imprinting distinct signatures on cosmological observables like the CMB anisotropies and large-scale structure. By comparing these theoretical predictions with observational data from cosmological probes, one can rigorously test and constrain different models of inflation~\cite{Lyth:2005fi,Martin:2013tda,Martin:2015dha,Planck:2018jri}. 

As mentioned earlier, initial conditions for quantum fluctuations are imposed when their associated length scales are smaller than the Hubble horizon size. During this phase, the background spatial geometry appears flat at short distances, and quantum fluctuations are expected to begin in a minimum energy state. This motivates the standard choice of the Bunch-Davies vacuum as the simplest and most physically plausible initial state for primordial quantum fluctuations. As inflation proceeds, these modes exit the Hubble horizon, and their quantum state evolves dynamically into a two-mode squeezed state due to the rapid cosmological expansion. This squeezing mechanism during inflation has been extensively studied through various quantum information measures, including Bell inequalities, entanglement entropy, entanglement negativity, and quantum discord, assuming the Bunch-Davies vacuum as given in~\cite{Albrecht:1992kf,Kanno:2017dci,Martin:2015qta,brahma2020entanglement,Boutivas:2023ksg}. 

However, the Bunch–Davies vacuum is not the unique choice for the initial state of primordial perturbations. Several studies have explored the cosmological consequences of deviations from the Bunch–Davies vacuum (see, for example, Refs.~\cite{Sriramkumar:2004pj,Kundu:2011sg,Holman:2007na}). Among the alternatives, the coherent state as an initial state is of particular interest. A defining feature of perturbations in a coherent state is the presence of a non-vanishing one-point correlation function \cite{Kundu:2011sg}—a property absent in the vacuum state. Moreover, the presence of a non-zero one-point correlation function generically leads to a breakdown of statistical homogeneity and isotropy of the primordial perturbations, giving rise to potentially testable cosmological signatures~\cite{Kundu:2011sg, Kundu:2013gha, Ragavendra:2024qpj, Mukherjee:2025dcv}. 

Motivated by these considerations, in this work we take the coherent state as the initial state for inflationary perturbations and study the Bell inequality violations in this context. As mentioned above, Bell inequality violations have been studied previously in the context of inflation when the initial state is the Bunch–Davies vacuum (see Refs.~\cite{Martin:2017zxs, Martin:2016tbd}). However, the possibility of Bell inequality violations when the perturbations begin in a coherent state—characterized by a non-zero one-point correlation—has not yet been explored. In this work, we take a step in this direction.

The organization of this article is as follows. In \ref{sec:2}, we provide a brief review of the theory of inflationary perturbations. In \ref{sec:3}, we canonically quantize these perturbations and express the time-evolution operator in terms of squeezing and rotation operators. We then consider a coherent state, rather than the commonly adopted Bunch–Davies vacuum, as the initial state and study the action of the time-evolution operator on it. Subsequently, we compute the one-point and two-point correlation functions associated with the quadrature variable and its conjugate momentum. In \ref{sec:7}, we present a brief introduction to Bell inequality and explicitly construct the Bell operator using the Gour–Khanna–Mann–Revzen (GKMR) pseudo-spin operators. We then analytically evaluate the expectation value of the Bell operator with respect to the squeezed coherent state. Finally, we specialize to the de Sitter limit and plot the expectation value of the Bell operator as a function of the squeezing parameter. The main conclusions of our work are summarized in \ref{sec:8}.

Throughout this article, we use the following metric signature: $(-,+,+,+)$. For background cosmology, we use spatially flat Friedmann-Lemaitre-Robertson-Walker (FLRW) metric which is written as, $ds^{2}=a^{2} \big(-d\eta^{2} + \delta_{ij}dx^{i}dx^{j} \big)$. In this article, over-dot denotes the derivative with respect to cosmic time $t$ and over-prime denotes the derivative with respect to conformal time $\eta$. Also, the Hubble parameter $H$ is defined as~$H=\dot{a}/a$.

\color{black}
\section{Inflationary cosmological perturbations}\label{sec:2}

The evolution of inflationary perturbations can be studied using linear cosmological perturbation theory, where scalar, tensor, and vector perturbations evolve independently. In this work, we focus specifically on the evolution of scalar perturbations and discuss their dynamics in detail. 

During the inflationary epoch, scalar perturbations are typically described in terms of a single dynamical degree of freedom, the Mukhanov–Sasaki variable $v(x, t)$. The second–order action governing these perturbations can be expressed as 
\begin{eqnarray}\label{action_2}
    ^{(2)}S &=& \frac{1}{2} \int d^{4}x \bigg[(v')^{2} - \delta^{ij}\partial_{i}v\partial_{j}v + \frac{z''}{z}v^{2} \bigg]\, ,  
\end{eqnarray}
where $z = a \, \dot{\overline{\phi}}/H $, \, with $\overline{\phi}$ being the scalar field which drives the inflation. Instead of working in real space, it is often more convenient to switch into Fourier space on a constant time slice since all Fourier modes evolve independently. Moreover, to ensure that the Fourier transform retains the reality of the original position-space Mukhanov–Sasaki variable $v(x, t)$, the Fourier modes must satisfy the condition $v^{*}_{\mathbf{k}}=v_{-\mathbf{k}}$.  This constraint implies that $v_{\mathbf{k}}$ is, in general, complex. By taking the Fourier transform of all the terms in the second-order action \ref{action_2}, simplifies to the following form   
\begin{equation}\label{Fourier}
    ^{(2)}S \, = \, \frac{1}{2} \int d\eta \hspace{0.07cm} d^{3}k \bigg[v'_{\mathbf{k}}v'^{*}_{\mathbf{k}} - \bigg(k^{2}- \frac{z''}{z} \bigg)v_{\mathbf{k}}v^{*}_{\mathbf{k}} \bigg]\,. 
\end{equation}
It is well known that the addition of a total derivative term to the action does not affect the resulting equations of motion. Utilizing this freedom, one can add a total derivative term  $\frac{1}{2}(\frac{z'}{z} v_{\mathbf{k}}v^{*}_{\mathbf{k}})'$, to the action \ref{Fourier}  and obtain
\begin{equation}\label{Lagrangian}
    ^{(2)}S \, = \, \frac{1}{2} \int_{\mathbb{R}^{3+}} d\eta \hspace{0.07cm} d^{3}k \bigg[(v'_{\mathbf{k}}v'^{*}_{\mathbf{k}} + v'^{*}_{\mathbf{k}}v'_{\mathbf{k}}) - 2\frac{z'}{z}\big(v_{\mathbf{k}}v'^{*}_{\mathbf{k}} + v^{*}_{\mathbf{k}}v'_{\mathbf{k}} \big) + \bigg(\frac{z'^{2}}{z^{2}}-k^{2} \bigg) \big(v_{\mathbf{k}}v^{*}_{\mathbf{k}} + v^{*}_{\mathbf{k}}v_{\mathbf{k}} \big) \bigg]\,. 
\end{equation}
By varying the action, one obtains the equation of motion for the Mukhanov–Sasaki variable as
\begin{equation}\label{EOM}
    v''_{\mathbf{k}} + \bigg(k^{2}-\frac{z''}{z} \bigg)v_{\mathbf{k}} = 0 \,. 
\end{equation}
Interestingly, this equation takes the form of a harmonic oscillator with a time-dependent frequency $\omega(\eta) = \big(k^{2} - z''/z \big)^{1/2}$.

Our next task is to determine the Hamiltonian density associated with the perturbations. Using the standard Legendre transformation, the Hamiltonian can be found as 
\begin{eqnarray}\label{Hamiltonian_1} 
    H &=& \int d^{3}k \hspace{0.06cm} \bigg[p_{\mathbf{k}}p^{*}_{\mathbf{k}} + \frac{z'}{z} \big(p_{\mathbf{k}}v^{*}_{\mathbf{k}} + p^{*}_{\mathbf{k}}v_{\mathbf{k}} \big) + k^{2}v_{\mathbf{k}}v^{*}_{\mathbf{k}} \bigg]\,.   
\end{eqnarray}
Here $p_{\mathbf{k}}$ is the conjugate momentum and can be obtained as
\begin{eqnarray}
    p_{\mathbf{k}} &=& v'_{\mathbf{k}} - \frac{z'}{z}v_{\mathbf{k}}\,. 
\end{eqnarray}
As expected, the structure of this Hamiltonian \ref{Hamiltonian_1} resembles with that of a time-dependent harmonic oscillator.

\section{Quantization of perturbations}\label{sec:3}

So far, we have discussed the classical dynamics of cosmological perturbations. The next step is to quantize these perturbations, which involves promoting the classical Mukhanov–Sasaki variable $v_{\mathbf{k}}$ and its conjugate momentum $p_{\mathbf{k}}$ to quantum operators $\hat{v}_{\mathbf{k}}$ and $\hat{p}_{\mathbf{k}}$, respectively. These operators are then expressed in terms of creation and annihilation operators $\hat{a}^{\dagger}_{\mathbf{k}}$ and $\hat{a}_{\mathbf{k}}$, such that the canonical commutation relation~\cite{Mukhanov:1988jd}, 
\begin{equation}
[\hat{v}_{\mathbf{k}}, \hat{p}_{\mathbf{k}'}] \, =  \, i \, \delta(\mathbf{k} - \mathbf{k}')\, , 
\end{equation}
is preserved. A convenient decomposition that satisfies this requirement is given by:
\begin{eqnarray}\label{decomp}
    \hat{v}_{\mathbf{k}} &=& \frac{1}{\sqrt{2 k}} \left(\hat{a}_{\mathbf{k}} + \hat{a}^{\dagger}_{-\mathbf{k}} \right), \\ 
    \hat{p}_{\mathbf{k}} &=&- i\,\sqrt{\frac{k}{2} }\left(\hat{a}_{\mathbf{k}} - \hat{a}^{\dagger}_{-\mathbf{k}} \right)\,. 
\end{eqnarray}
This decomposition explicitly mixes the ladder operators $\hat{a}_{\mathbf{k}} $ and $ \hat{a}^{\dagger}_{-\mathbf{k}}$, corresponding to opposite momentum modes. This mixing is essential to ensure the correct Hermiticity condition for the field operator. Substituting this decomposition into the Hamiltonian expression \ref{Hamiltonian_1}, the quantized Hamiltonian takes the following form:
\begin{equation}\label{Hamiltonian}
    \hat{H} = \int d^{3}k \left[ \frac{k}{2} \left( \hat{a}_{\mathbf{k}} \hat{a}^{\dagger}_{\mathbf{k}} + \hat{a}_{-\mathbf{k}} \hat{a}^{\dagger}_{-\mathbf{k}} \right) - \frac{i}{2} \frac{z'}{z} \left( \hat{a}_{\mathbf{k}} \hat{a}_{-\mathbf{k}} - \hat{a}^{\dagger}_{-\mathbf{k}} \hat{a}^{\dagger}_{\mathbf{k}} \right) \right]\,.
\end{equation}
It is important to note that the first term in the above Hamiltonian represents a collection of free harmonic oscillators, each corresponding to a mode $\mathbf{k}$. In contrast, the second term describes a non-trivial interaction between the modes $\mathbf{k}$ and $-\mathbf{k}$. This interaction arises purely due to the time dependence of the background spacetime and is characterized by the coupling function $z'/z$. As such, the strength of this interaction is directly governed by the dynamics of the inflationary background.

Our next objective is to study the time evolution of the state associated with the Mukhanov–Sasaki variable. This is generally achieved by solving the time-dependent Schr\"{o}dinger equation with a suitable initial condition. An alternative but equivalent approach involves the use of the time evolution operator $\hat{U}(t_{0},t)$ which has the form 
\begin{equation}
\hat{U}(t_{0}, t) \hspace{0.15 cm} = \hspace{0.15 cm} \mathcal{T}\left\{\exp\left[-i\int^{t}_{t_{0}} dt'\hat{H}(t')/\hbar \right] \right\}\,. 
\end{equation} 
Here $\mathcal{T}$ denotes the time ordering of the respective operators. For the specific Hamiltonian given in \ref{Hamiltonian}, it turns out that the time evolution operator can be factorized into the squeezing $\hat{S}_{\mathbf{k}}(r_{k},\phi_{k})$ and rotation $\hat{R}_{\mathbf{k}}(\theta_{k})$ operator as given below~\cite{Albrecht:1992kf,Ma:1990llj,PhysRevA.39.1941} 
\begin{equation}\label{time_evolution}
    \hat{U}_{\mathbf{k}}(r_{k}, \theta_{k}, \phi_{k}) \hspace{0.15 cm} = \hspace{0.15 cm} \hat{S}_{\mathbf{k}}(r_{k},\phi_{k}) \hat{R}_{\mathbf{k}}(\theta_{k})\,. 
\end{equation}
In terms of ladder operators, the squeezing $\hat{S}_{\mathbf{k}}(r_{k},\phi_{k})$ and rotation $\hat{R}_{\mathbf{k}}(\theta_{k})$ operator can be expressed as 
\begin{eqnarray}\label{squeezing_rotation}
    \hat{S}_{\mathbf{k}}(r_{k}, \phi_{k}) &=& \exp \left\{r_{k} \left[e^{-2i\phi_{k}}\hat{a}_{\mathbf{k}}(\eta_{0}) \, \hat{a}_{-\mathbf{k}}(\eta_{0}) - e^{2i\phi_{k}}\hat{a}^{\dagger}_{\mathbf{k}}(\eta_{0}) \, \hat{a}^{\dagger}_{-\mathbf{k}}(\eta_{0}) \right]\right\}\, ,  \nonumber \\
    \hat{R}_{\mathbf{k}}(\theta_{k}) &=& \exp \left\{-i\theta_{k} \left[\hat{a}^{\dagger}_{\mathbf{k}}(\eta_{0}) \, \hat{a}_{\mathbf{k}}(\eta_{0}) + \hat{a}^{\dagger}_{-\mathbf{k}}(\eta_{0}) \, \hat{a}_{-\mathbf{k}}(\eta_{0}) \right] \right\}\,. 
\end{eqnarray}
Here $\eta_{0}$ refers to the initial conformal time when the Fourier modes are deep inside the horizon. The quantities $r_{k},\phi_{k}$ and $\theta_{k}$ correspond to the squeezing parameter, squeezing angle and rotation angle respectively, all of which evolve with time. For the equation of motions governing the squeezing parameters $r_{k},\phi_{k}$ and $\theta_{k}$, see \cite{Martin_2016}.

\subsection{Choice of initial state}

The Bunch-Davies vacuum is typically chosen as the initial state for inflationary perturbations. However, in this article, we explore a possible alternative—namely, the coherent state—as a candidate for the initial condition \cite{Kundu:2011sg}. A coherent state is defined as the eigenstate of the annihilation operator as follows,
\begin{equation}
    \hat{a}_{\mathbf{k}} \ket{\alpha_{\mathbf{k}}} \hspace{0.15 cm} = \hspace{0.15 cm} \alpha_{\mathbf{k}} \ket{\alpha_{\mathbf{k}}} . 
\end{equation}
Due to the non-Hermitian nature of annihilation operator $\hat{a}_{\mathbf{k}}$, coherent states $\ket{\alpha_{\mathbf{k}}}$ do not form a mutually orthogonal set of basis vectors and their corresponding eigenvalues are not necessarily real. As a result, one has to deal with complex eigen spectrum in the context of coherent state. Alternatively, coherent states can be constructed using the operator $\hat{D}(\alpha_{\mathbf{k}})$ as given below,   
\begin{equation}\label{Displacement}
    \ket{\alpha_{\mathbf{k}}} \hspace{0.15 cm} = \hspace{0.15 cm} \hat{D}(\alpha_{\mathbf{k}}) \ket{0_{\mathbf{k}}} \,, 
\end{equation}
where the displacement operator is defined as, $\hat{D}(\alpha_{\mathbf{k}}) \, = \, \exp\big(\alpha_{\mathbf{k}} \, \hat{a}^{\dagger}_{\mathbf{k}} - \alpha^{*}_{\mathbf{k}} \, \hat{a}_{\mathbf{k}} \big)$. 

Let us briefly describe the action of the displacement operator, $\hat{D}(\alpha_{\mathbf{k}})$, on the wave function of a Gaussian state. As the name suggests, the operator $\hat{D}(\alpha_{\mathbf{k}})$ effectively shifts the peak of the Gaussian wave packet from the origin to a new point in the given coordinate. This shift is precisely what manifests itself as non-vanishing one-point correlation functions, which is absent in the case of vacuum state. In the context of inflationary perturbations described by the Mukhanov–Sasaki variable $\hat{v}_{\mathbf{k}}$, we obtain 
\begin{equation}\label{Mukhanov_correlation}
    \left\langle\alpha_{\mathbf{k}} \right| \hat{v}_{\mathbf{k}} \left|\alpha_{\mathbf{k}} \right\rangle = \frac{1}{\sqrt{2 k}} \big(\alpha_{\mathbf{k}} + \alpha^{*}_{-\mathbf{k}} \big) \,,\quad \left\langle\alpha_{\mathbf{k}} \right| \hat{p}_{\mathbf{k}} \left|\alpha_{\mathbf{k}} \right\rangle = -i\,\sqrt{\frac{k}{2} }\big(\alpha_{\mathbf{k}} - \alpha^{*}_{-\mathbf{k}} \big) \,. 
\end{equation}
As evident from the \ref{Mukhanov_correlation}, the one-point correlation functions turn out to be complex because the Mukhanov-Sasaki variable $\hat{v}_{\mathbf{k}}$ and its conjugate momentum $\hat{p}_{\mathbf{k}}$ are non-Hermitian operators. Nevertheless, by choosing a suitable combination of $\hat{v}_{\mathbf{k}}$ and $\hat{p}_{\mathbf{k}}$, one can obtain such one-point correlations that are entirely real. For detailed expressions of such one-point correlations, see the \ref{sec:3.3}. Moreover, the coherent-state parameters $\alpha_{\mathbf{k}}$ and $\alpha_{-\mathbf{k}}$ play a central role for the presence of non-vanishing one-point correlation function.

Performing the action of the displacement operator on the vacuum state, as given in \ref{Displacement}, leads to the following expression of coherent state  
\begin{equation}
    \ket{\alpha_{\mathbf{k}}} \hspace{0.15 cm} = \hspace{0.15 cm} \exp\bigg(-\frac{\abs{\alpha_{\mathbf{k}}}^{2}}{2} \bigg) \sum_{n=0}^{\infty} \frac{(\alpha_{\mathbf{k}})^{n}}{\sqrt{n!}} \ket{n} \,. 
\end{equation}
Here $\ket{n}$ is the eigenstate of the Hamiltonian which belongs to a system of harmonic oscillators. For more details, see \cite{Bagchi_2020,Glauber:1963tx}.

In the context of inflationary perturbations, the Hilbert space associated with Mukhanov-Sasaki variable has been split into two sectors corresponding to Fourier modes $\mathbf{k}$ and $\mathbf{-k}$. Therefore, we define the coherent state of this bi-partied Hilbert space in the following manner:
\begin{equation}\label{2_coherent}
    \ket{\alpha_{\mathbf{k}}\alpha_{-\mathbf{k}}} \hspace{0.15 cm} = \hspace{0.15 cm} \ket{\alpha_{\mathbf{k}}} \otimes \ket{\alpha_{-\mathbf{k}}} \,. 
\end{equation} 
It is usually referred to as two mode coherent state which has the intriguing property of minimizing the Heisenberg uncertainty principle \cite{Bagchi_2020}. This property serves as a motivation for its choice as an initial state. Moreover, in terms of the wave function of coherent state $\ket{\alpha_{\mathbf{k}} \alpha_{-\mathbf{k}}}$, one can explicitly verify its displaced Gaussian nature \cite{Hartley:1982ygx,choi2004coherent}.

\subsection{Squeezed coherent state}

As we have discussed, during inflation, Fourier modes of cosmological perturbations get stretched and eventually exit the Hubble radius. While the classical dynamics of cosmological perturbations can be evaluated using the Mukhanov–Sasaki equation, the evolution of the quantum state associated with these perturbations is best understood using the squeezing formalism, as introduced earlier.

It is well known that when the perturbations are initialized in the Bunch–Davies vacuum state, their evolution under inflationary dynamics leads to a squeezed vacuum state, denoted by $\ket{r,0}$ \cite{Martin:2015qta}. However, if the initial state is instead a coherent state, the system evolves into a squeezed coherent state, denoted by $\ket{r,\alpha}$. This state can be written as,  
\begin{eqnarray}\label{squeezed coherent}
    \ket{r,\alpha} &=& \hat{U}_{\mathbf{k}}(r_{k}, \theta_{k}, \phi_{k}) \hspace{0.07 cm} \ket{\alpha_{\mathbf{k}},\alpha_{-\mathbf{k}}} \hspace{0.4 cm} = \hspace{0.2 cm} \hat{S}_{\mathbf{k}}(r_{k},\phi_{k}) \hat{R}_{\mathbf{k}}(\theta_{k}) \hspace{0.07 cm} \hat{D}(\alpha_{\mathbf{k}}) \hat{D}(\alpha_{-\mathbf{k}}) \ket{0_{\mathbf{k}}, 0_{-\mathbf{k}}}\, . 
\end{eqnarray}
Here, the displacement operators $\hat{D}(\alpha_{\mathbf{k}})$ introduce an initial shift in the peak of the Gaussian wave function, generating non-vanishing one-point correlation functions, while the squeezing and rotation operators $\hat{S}_{\mathbf{k}}(r_{k},\phi_{k})$ and $\hat{R}_{\mathbf{k}}(\theta_{k})$ encode the characteristic mode squeezing produced during inflation.

To express this state $\ket{r,\alpha}$ in the number basis $\ket{n_{\mathbf{k}}, n_{-\mathbf{k}}}$, we need to analyze the action of the squeezing and rotation operators on the initial coherent state defined in~\ref{2_coherent}. The squeezing operator $\hat{S}_{\mathbf{k}}(r_{k},\phi_{k})$ can be factorized into three exponentials involving the ladder operators $\hat{a}_{\mathbf{k}}$ and $\hat{a}^{\dagger}_{-\mathbf{k}}$. This factorization is given by \cite{10.1093/acprof:oso/9780198563617.001.0001,ZAYED2005967} 
\begin{equation}\label{squeezing_fact}
    \hat{S}_{\mathbf{k}}(r_{k},\phi_{k}) = \exp\left(e^{-2i\phi_{k}}\tanh{r_{k}} \hspace{0.1cm} \hat{a}^{\dagger}_{-\mathbf{k}}\hat{a}^{\dagger}_{\mathbf{k}} \right) \exp[-\ln{\big(\cosh{r_{k}} \big)} \hspace{0.1cm} \big( \hat{a}^{\dagger}_{\mathbf{k}}\hat{a}_{\mathbf{k}} + \hat{a}_{-\mathbf{k}} \hat{a}^{\dagger}_{-\mathbf{k}} \big)] \exp\left(-e^{2i\phi_{k}}\tanh{r_{k}} \hspace{0.1cm} \hat{a}_{-\mathbf{k}}\hat{a}_{\mathbf{k}} \right)\,. 
\end{equation}
For the detailed derivation, see Appendix~\ref{appendix:C}. The action of the last two exponentials in~\ref{squeezing_fact} on the coherent state is straightforward, as the coherent state is an eigenstate of the corresponding annihilation operators. However, the first term in~\ref{squeezing_fact} is mainly responsible for the entangled nature of the two-mode squeezed coherent state. After performing the necessary calculations, the state in~\ref{squeezed coherent} can be expressed in the number basis as 
\begin{equation}\label{SC_number}
    \ket{r,\alpha} \hspace{0.15 cm} = \hspace{0.15 cm} A_{0}(r_{k}, \theta_{k}, \phi_{k}) \sum_{n_{k}}\sum_{m_{-k}} \sum_{p} C_{n_{k}m_{-k}p} \ket{(n+p)_{k};(m+p)_{-k}} \,. 
\end{equation}
The definition of the coefficient $C_{n_{k}m_{-k}p}$ and the normalization factor $A_{0}(r_{k},\theta_{k},\phi_{k})$ are given below, 
\begin{eqnarray}
    C_{n_{k}m_{-k}p} &=& \frac{(\alpha_{\mathbf{k}})^{n_{k}}}{n_{k}!} \hspace{0.07cm} \frac{(\alpha_{-\mathbf{k}})^{m_{-k}}}{m_{-k}!} \hspace{0.07cm} \frac{(-e^{2i\phi_{k}} \tanh{r_{k}})^{p}}{p!} \big[e^{-i\theta_{k}-\ln{(\cosh{r}_{k})}} \big]^{n_{k}+m_{-k}} \sqrt{(n_{k}+p)!(m_{-k}+p)!} \, , \\
    A_{0}(r_{k},\theta_{k},\phi_{k}) &=& \exp[-i\theta_{k} - \ln{(\cosh{r_{k}})} + \alpha_{\mathbf{k}}\alpha_{-\mathbf{k}} e^{-2i(\theta_{k}+\phi_{k})} \tanh{r_{k}} ] \hspace{0.08 cm} \exp[-\frac{1}{2} \left(\abs{\alpha_{\mathbf{k}}}^{2}+\abs{\alpha_{-\mathbf{k}}}^{2} \right)]\,. 
\end{eqnarray}
In the limit $(\alpha_{\mathbf{k}},\alpha_{-\mathbf{k}})\to 0$, it can be shown that the above state \ref{SC_number} exactly coincides with that of a two mode squeezed vacuum state, as expected \cite{Martin:2015qta}.

\subsection{Correlation functions}\label{sec:3.3}

From the decomposition given in~\ref{decomp}, it is evident that the Mukhanov-Sasaki variable is not Hermitian, as $\hat{v}_{\mathbf{k}}^\dagger = \hat{v}_{-\mathbf{k}}$. As a result, the associated correlation functions generally turn out to be complex. However, physical observables must correspond to real quantities, and hence it is necessary to reformulate the problem using an alternative set of conjugate variables that satisfy Hermiticity. To this end, let us define the quadrature operators $\hat{q}_{\mathbf{k}}$ and $\hat{\pi}_{\mathbf{k}}$, as specific linear combinations of the Mukhanov-Sasaki variable $\hat{v}_{\mathbf{k}}$ and its conjugate momentum $\hat{p}_{\mathbf{k}}$ in the following manner:
\begin{eqnarray}\label{Mukhanov_Quadrature_0}
    \sqrt{k} \, \hat{v}_{\mathbf{k}} &=& \frac{1}{2} \left[\big(\hat{q}_{\mathbf{k}}+\hat{q}_{\mathbf{-k}} \big) + i \big(\hat{\pi}_{\mathbf{k}}-\hat{\pi}_{-\mathbf{k}} \big) \right] \,, \\ 
    \frac{\hat{p}_{\mathbf{k}}}{\sqrt{k}} &=& \frac{1}{2\,i} \left[\big(\hat{q}_{\mathbf{k}}-\hat{q}_{\mathbf{-k}} \big) + i \big(\hat{\pi}_{\mathbf{k}}+\hat{\pi}_{-\mathbf{k}} \big) \right] \,. 
\end{eqnarray}
This transformation ensures that the quadrature variables $\hat{q}_{\mathbf{k}}$ and $\hat{\pi}_{\mathbf{k}}$ are dimensionless and they satisfy the condition, i.e. $\hat{q}^\dagger_{\mathbf{k}} = \hat{q}_{\mathbf{k}}$ and $\hat{\pi}^\dagger_{\mathbf{k}} = \hat{\pi}_{\mathbf{k}}$, thus restoring the Hermitian nature of the field operators and allowing us to compute real-valued correlation functions. The exact decomposition of quadrature variables in terms of ladder operators are given below, 
\begin{eqnarray}\label{Quadrature_decomp}
    \hat{q}_{\mathbf{k}}\, = \,\frac{1}{2} \left[\hat{a}_{\mathbf{k}}(\eta_{0}) + \hat{a}^{\dagger}_{\mathbf{k}}(\eta_{0}) \right]\,, \hspace{0.4cm}  \hspace{0.4cm} \hat{\pi}_{\mathbf{k}} \, = \, \frac{1}{2\,i} \left[\hat{a}_{\mathbf{k}}(\eta_{0}) - \hat{a}^{\dagger}_{\mathbf{k}}(\eta_{0}) \right] \,. 
\end{eqnarray}
From the decomposition given above, it follows that the quadrature variables are Hermitian operators i.e. $\hat{q}_{\mathbf{k}}=\hat{q}^{\dagger}_{\mathbf{k}}$ and $\hat{\pi}_{\mathbf{k}}=\hat{\pi}^{\dagger}_{\mathbf{k}}$. Unlike the Mukhanov-Sasaki variable, this Hermitian nature of quadrature variables is primarily responsible for non-mixed nature of Fourier modes $\mathbf{k}$ and $\mathbf{-k}$ in the \ref{Quadrature_decomp}.

It is often useful to understand the physical meaning of the squeezed coherent state by expressing it in the $(q_{\mathbf{k}}, q_{-\mathbf{k}})$ basis. A detailed derivation of the corresponding wavefunction can be found in the Appendix~\ref{appendix:B}. For convenience, the final expression is given below:

\begin{equation}\label{Wave function}
    \psi_{r}(q_{\mathbf{k}},q_{-\mathbf{k}}) \, = \, N_{0} \exp[-A_{0}\big(q_{\mathbf{k}}-C_{0} \big)^{2} - A_{0}\big(q_{-\mathbf{k}}-D_{0} \big)^{2} + B_{0} \hspace{0.06cm} q_{\mathbf{k}}q_{-\mathbf{k}} ] \,. 
\end{equation}
Interestingly, this wave function takes the form of a displaced gaussian function whose peak value is shifted from the origin of the chosen coordinate system. The coefficients $A_{0}, B_{0}, C_{0}, D_{0}$ and the normalization factor $N_{0}$, expressed in terms of the squeezing parameters $r_{k}, \theta_{k}$ and $\phi_{k}$, are given below    
\begin{eqnarray}
    A_{0} &=& \frac{1}{2} \bigg(\frac{1+e^{-4i\phi_{k}}\tanh^{2}{r_{k}}}{1-e^{-4i\phi_{k}}\tanh^{2}{r_{k}}} \bigg) \,, \\ 
    B_{0} &=& \bigg(\frac{2 \, e^{-2i\phi_{k}}\tanh{r_{k}}}{1-e^{-4i\phi_{k}}\tanh^{2}{r_{k}}} \bigg)\, , \\ 
    C_{0} &=& \frac{1}{\big(1-e^{-4i\phi_{k}}\tanh^{2}{r_{k}} \big)} \big[\big(\beta_{\mathbf{k}} + e^{-4i\phi_{k}}\tanh^{2}{r_{k}} \hspace{0.07cm} \beta^{*}_{\mathbf{k}} \big) - 2e^{-2i\phi_{k}}\tanh{r_{k}} \Re\big(\gamma_{\mathbf{k}} \big) \big] \,, \\ 
    D_{0} &=& \frac{1}{\big(1-e^{-4i\phi_{k}}\tanh^{2}{r_{k}} \big)} \big[\big(\gamma_{\mathbf{k}} + e^{-4i\phi_{k}}\tanh^{2}{r_{k}} \hspace{0.07cm} \gamma^{*}_{\mathbf{k}} \big) - 2e^{-2i\phi_{k}}\tanh{r_{k}} \Re\big(\beta_{\mathbf{k}} \big) \big] \, , \\ 
    N_{0} &=& \frac{\sech{r_{k}}}{\sqrt{\pi (1-e^{-4i\phi_{k}}\tanh^{2}{r_{k}})}} \exp\left\{-4i\Im\big(A_{0} \big) \left[\Re\big(\beta_{\mathbf{k}} \big)^{2} + \Re\big(\gamma_{\mathbf{k}} \big)^{2} \right] + 4i\Im\big(B_{0} \big) \,  \Re\big(\beta_{\mathbf{k}} \big)\Re\big(\gamma_{\mathbf{k}} \big) \right\} \nonumber \\ 
    && \cross \, \exp\left\{A_{0} \left(C_{0}^{2} + D_{0}^{2} \right) - 2i \, \left[\Re\big(\beta_{\mathbf{k}} \big)\Im\big(\beta_{\mathbf{k}} \big) + \Re\big(\gamma_{\mathbf{k}} \big) \Im\big(\gamma_{\mathbf{k}} \big) \right] - \frac{1}{2} \left(g_{1}x^{2}_{0} + g_{2}y^{2}_{0} \right) \right\} \, \label{Normalization_sc} .   
\end{eqnarray}
The quantities $\beta_{\mathbf{k}}$ and $\gamma_{\mathbf{k}}$ are expressed in the following manner
\begin{eqnarray}\label{beta_gamma}
    \beta_{\mathbf{k}} &=& \big(e^{-i\theta_{k}}\cosh{r_{k}} \hspace{0.1cm} \alpha_{\mathbf{k}} - \hspace{0.1cm} e^{i(\theta_{k}+2\phi_{k})}\sinh{r_{k}} \hspace{0.1cm} \alpha^{*}_{-\mathbf{k}} \big)\, ,  \\ 
    \gamma_{\mathbf{k}} &=& \big(e^{-i\theta_{k}}\cosh{r_{k}} \hspace{0.1cm} \alpha_{-\mathbf{k}} - \hspace{0.1cm} e^{i(\theta_{k}+2\phi_{k})}\sinh{r_{k}} \hspace{0.1cm} \alpha^{*}_{\mathbf{k}} \big)\,. 
\end{eqnarray}
The appearance of several unknown quantities $g_{1}, g_{2}, x_{0}$ and $y_{0}$ without any prior definition, has been found in the normalization factor $N_{0}$, as given in \ref{Normalization_sc}. Their explicit expressions are provided in the \ref{Subsection-spin-xx}, specifically through the \ref{def-g_1}~-~\ref{def-y_0}.

Before proceeding, it is important to highlight some key properties of the wave function given above. One immediate consequence of choosing a coherent state as the initial state is the displacement of peak of the given wave function from the origin to a non-zero coordinate point. Another noteworthy feature of the wave function in the quadrature basis is the presence of cross terms like~$q_{\mathbf{k}} q_{\mathbf{-k}}$, which signals entanglement between the Fourier modes $\mathbf{k}$ and $\mathbf{-k}$. It is well established that the degree of entanglement increases as inflation progresses. This entanglement can be removed by transforming to a properly diagonalized basis, in which the wave function appears as a product state. In this new basis, the entanglement manifests as squeezing—that is, a redistribution of uncertainty between the conjugate variables. As a result, the uncertainty becomes compressed along one direction (for example, the field variable $v_{\mathbf{k}}$) and stretched along the conjugate momentum direction $p_{\mathbf{k}}$. This transformation is captured by the action of the squeezing operator, \ref{squeezing_fact}, which is characterized by the squeezing parameters. The extent of squeezing is quantified by the squeezing parameter $r_{k}$. Typically, during inflation, the quantum state becomes highly squeezed, indicated by a large value of $r_{k}$ (see detailed discussions in\cite{Albrecht:1992kf, Martin:2004um, Martin:2019wta}). 

Using the above decomposition \ref{Quadrature_decomp}, one can compute the one point and two point correlation function of quadrature variables, as described below 
\begin{eqnarray}\label{Correlation_quad_Schr}
    \left\langle r,\alpha \right| \hat{q}_{\mathbf{k}} \left| r,\alpha \right\rangle &=&  \Re\big(\beta_{\mathbf{k}} \big) \,, \\
    \left\langle r,\alpha \right| \hat{\pi}_{\mathbf{k}} \left| r,\alpha \right\rangle &=& \Im\big(\beta_{\mathbf{k}} \big) \, , \\
    \left\langle r,\alpha \right| \hat{q}_{\mathbf{k}}\hat{q}_{\mathbf{k}} \left| r,\alpha \right\rangle
    &=& \left[\Re\big(\beta_{\mathbf{k}} \big) \right]^2 + \frac{1}{4} \cosh{2r_{k}} \, , \\ 
    \left\langle r,\alpha \right| \hat{\pi}_{\mathbf{k}}\hat{\pi}_{\mathbf{k}} \left| r,\alpha \right\rangle &=& \left[\Im\big(\beta_{\mathbf{k}} \big) \right]^{2} + \frac{1}{4} \cosh{2r_{k}} \, , \\ 
    \left\langle r,\alpha \right| \hat{q}_{\mathbf{k}}\hat{\pi}_{\mathbf{k}} \left| r,\alpha \right\rangle &=& \Re\big(\beta_{\mathbf{k}} \big) \Im\big(\beta_{\mathbf{k}} \big) - \frac{1}{4i}\, ,\\
    \left\langle r,\alpha \right| \hat{q}_{\mathbf{k}}\hat{q}_{\mathbf{-k}} \left| r,\alpha \right\rangle  &=& \Re\big(\beta_{\mathbf{k}} \big) \Re\big(\gamma_{\mathbf{k}} \big) - \frac{1}{4} \cos{2\phi_{k}} \, \sinh{2r_{k}} \,, \\ 
    \left\langle r,\alpha \right| \hat{\pi}_{\mathbf{k}}\hat{\pi}_{-\mathbf{k}} \left| r,\alpha \right\rangle &=& \Im\big(\beta_{\mathbf{k}} \big) \Im\big(\gamma_{\mathbf{k}} \big) + \frac{1}{4} \cos{2\phi_{k}} \, \sinh{2r_{k}} \, , \\ 
    \left\langle r,\alpha \right| \hat{q}_{\mathbf{k}}\hat{\pi}_{-\mathbf{k}} \left| r,\alpha \right\rangle &=& \Re\big(\beta_{\mathbf{k}} \big) \Im\big(\gamma_{\mathbf{k}} \big) - \frac{1}{4} \sinh{2r_{k}} \, \sin{2\phi_{k}} \,.\label{Correlation_quad}
\end{eqnarray}
All the correlation functions of the quadrature variables given in \ref{Correlation_quad_Schr}~-~\ref{Correlation_quad} explicitly depend on the coherent state parameters $\alpha_{\mathbf{k}}$ and $\alpha_{-\mathbf{k}}$. Moreover, it is easy to see that, in the limit $(\alpha_{\mathbf{k}}, \alpha_{-\mathbf{k}}) \to 0$, the results in \ref{Correlation_quad_Schr}~-~\ref{Correlation_quad} exactly reduce to those of the squeezed vacuum state~\cite{Martin_2016}.

\color{black}
\section{Bell Inequality violation}\label{sec:7}

As mentioned in the introduction, inflation leads to the production of entangled particle pairs, a phenomenon that is inherently quantum and lacks any classical analogue. Therefore, any measure that can quantify the quantum entanglement of primordial perturbations serves as a potential indicator of their quantum nature. One such measure is the violation of Bell inequalities.

Quantum entanglement inherently implies the non-local character of quantum mechanics — a property where the state of one particle in an entangled pair can be instantaneously influenced by measurements performed on its partner, regardless of the distance separating them, even beyond the light cone. In an effort to explain such puzzling features, several scientists proposed local hidden variable theories, which attempt to reproduce the probabilistic nature of quantum mechanics using deterministic but unobservable parameters.

However, John S. Bell formulated an inequality — now known as the Bell inequality — which provides a testable criterion for distinguishing quantum mechanics from any such local hidden variable theory \cite{Bell:1964kc,bell2004speakable}. In essence, Bell inequalities place bounds on the correlations that can be explained by local hidden variables. Quantum mechanics, on the other hand, predicts correlations that violate these bounds, thus ruling out the viability of local hidden variable models as a complete description of nature.

In practical settings, a more refined version of Bell's original inequality, known as the Clauser-Horne-Shimony-Holt (CHSH) inequality \cite{Clauser:1969ny}, is widely used to test for quantum non-locality. Violation of the CHSH inequality in an experiment confirms the presence of quantum entanglement and the fundamentally non-local nature of quantum mechanics.

In cosmology, we typically encounter physical observables which are continuous in space and time. This poses a challenge when attempting to directly apply Bell-type inequalities, which are traditionally formulated for discrete systems like spin-1/2 particles. However, recent advancements have demonstrated how pseudo-spin operators can be systematically constructed from continuous variables, such as position or momentum operators. This approach effectively maps the continuous-variable system to a discrete two-level system, enabling the application of CHSH-type Bell inequalities in a meaningful way.

\subsection{Bell operator in terms of GKMR pseudo-spin operators}
 
To examine the Bell-CHSH inequality for a  continuous variable quantum system, one has to construct a set of operators $\hat{S}_{x},\hat{S}_{y}$ and $\hat{S}_{z}$ which satisfy the same $SU(2)$ Lie algebra as that of Pauli spin matrices. This kind of operators are known as pseudo-spin operators. 

It is important to mention that the choice of pseudo-spin operators is not unique, and different constructions exist for continuous-variable quantum systems, including the Banaszek–Wodkiewicz (BW), Gour-Khanna-Mann-Revzen (GKMR) and Larsson prescriptions (see Ref.~\cite{Martin:2017zxs}). For the purposes of this work, we adopt the GKMR prescription \cite{GOUR2004415, Revzen:2004mzw}. 

According to GKMR approach, the pseudo-spin operators are defined as follows 
\begin{eqnarray}\label{pseudo_spin}
    \hat{S}_{x}(\mathbf{k}) &=& \int^{\infty}_{0} dq_{\mathbf{k}} \left(\ket{O_{\mathbf{k}}}\bra{E_{\mathbf{k}}} + \ket{E_{\mathbf{k}}}\bra{O_{\mathbf{k}}} \right) \,, \\ 
    \hat{S}_{y}(\mathbf{k}) &=& i \int^{\infty}_{0} dq_{\mathbf{k}} \left(\ket{O_{\mathbf{k}}}\bra{E_{\mathbf{k}}} - \ket{E_{\mathbf{k}}}\bra{O_{\mathbf{k}}} \right) \,, \label{pseudo-spin_20} \\ 
    \hat{S}_{z}(\mathbf{k}) &=& - \int^{\infty}_{0} dq_{\mathbf{k}} \left(\ket{E_{\mathbf{k}}}\bra{E_{\mathbf{k}}} - \ket{O_{\mathbf{k}}}\bra{O_{\mathbf{k}}} \right) \,. \label{pseudo-spin_30}
\end{eqnarray}
The basis kets $\ket{E_{\mathbf{k}}}$ and $\ket{O_{\mathbf{k}}}$ are defined as a linear combination of the eigenstates of quadrature operator $\ket{q_{\mathbf{k}}}$ and $\ket{-q_{\mathbf{k}}}$, which satisfy $\hat{q}_{\mathbf{k}}\ket{q_{\mathbf{k}}} = q_{\mathbf{k}}\ket{q_{\mathbf{k}}}$. The explicit forms of these basis states are given by 
\begin{eqnarray}
    \ket{E_{\mathbf{k}}} &=& \frac{1}{\sqrt{2}} \left(\ket{q_{\mathbf{k}}} + \ket{-q_{\mathbf{k}}} \right) \,, \\ 
    \ket{O_{\mathbf{k}}} &=& \frac{1}{\sqrt{2}} \left(\ket{q_{\mathbf{k}}} - \ket{-q_{\mathbf{k}}} \right) \,. 
\end{eqnarray}
The pseudo-spin operators in \ref{pseudo_spin}~-~\ref{pseudo-spin_30} are actually required to construct the Bell operator $\hat{\mathcal{B}}$ whose expectation value has to be evaluated for finding the Bell-CHSH inequality. The Bell operator is usually constructed in the following manner  
\begin{equation}\label{Bell_operator}
    \hat{\mathcal{B}} = \big(\hat{\vec{S}}_{\mathbf{k}}\cdot \vec{n}_{1}\otimes\hat{\vec{S}}_{-\mathbf{k}}\cdot \vec{n}_{2} \big) + \big(\hat{\vec{S}}_{\mathbf{k}}\cdot \vec{n}_{1}\otimes\hat{\vec{S}}_{-\mathbf{k}}\cdot \vec{n}^{'}_{2} \big) + \big(\hat{\vec{S}}_{\mathbf{k}}\cdot \vec{n}^{'}_{1}\otimes\hat{\vec{S}}_{-\mathbf{k}}\cdot \vec{n}_{2} \big) - \big(\hat{\vec{S}}_{\mathbf{k}}\cdot \vec{n}^{'}_{1}\otimes\hat{\vec{S}}_{-\mathbf{k}}\cdot \vec{n}^{'}_{2} \big) \,. 
\end{equation}
Here $\vec{n}_{1}, \vec{n}_{2}, \vec{n}^{'}_{1}$ and $\vec{n}^{'}_{2}$ are the unit vectors belonging to the 3D spin space. For simplicity, let us choose the azimuthal angle of these unit vectors $\vec{n}_{i}$, expressed in the spherical polar coordinate, to be equal to zero. Then in the spherical polar coordinate, the unit vectors can be written as,  
\begin{eqnarray}
    \vec{n}_{1} &=& (\sin{\theta_{1}} , 0 , \cos{\theta_{1}})\,, \hspace{0.3cm} \vec{n}^{'}_{1} = (\sin{\overline{\theta}_{1}},0,\cos{\overline{\theta}_{1}}) \, ,  \\ 
    \vec{n}_{2} &=& (\sin{\theta_{2}} , 0 , \cos{\theta_{2}}) \,, \hspace{0.3cm} \vec{n}^{'}_{2} = (\sin{\overline{\theta}_{2}},0,\cos{\overline{\theta}_{2}}) \, . 
\end{eqnarray}
Our next task is to find out the expectation value of Bell operator \ref{Bell_operator} with respect to the squeezed coherent state $\ket{r,\alpha}$ defined earlier in \ref{squeezed coherent}. To proceed systematically, we begin by calculating the first term in the Bell operator, denoted by $E(\theta_{1},\theta_{2})$ as
\begin{eqnarray}\label{spin_one}
    E(\theta_{1},\theta_{2}) &=& \left\langle r, \alpha \right | \hat{\vec{S}}_{\mathbf{k}}\cdot \vec{n}_{1}\otimes\hat{\vec{S}}_{-\mathbf{k}}\cdot \vec{n}_{2} \left | r, \alpha \right\rangle \nonumber \\
    &=&\sin{\theta_{1}}\sin{\theta_{2}} \, \boldsymbol{J_{1}(r,\alpha)} + \cos{\theta_{1}}\cos{\theta_{2}} \, \boldsymbol{J_{0}(r,\alpha)} + \sin{\theta_{1}}\cos{\theta_{2}} \, \boldsymbol{J_{2}(r,\alpha)} + \cos{\theta_{1}}\sin{\theta_{2}} \, \boldsymbol{J_{3}(r,\alpha)} \, .
\end{eqnarray}
Here the quantities $\boldsymbol{J_{i}(r,\alpha)}$ are defined as
\begin{eqnarray}
    \boldsymbol{J_{0}(r,\alpha)} &=& \left\langle r,\alpha \right|  \hat{S}_{z}(\mathbf{k})\hat{S}_{z}(-\mathbf{k}) \left| r,\alpha \right\rangle \,, \\ 
    \boldsymbol{J_{1}(r,\alpha)} &=& \left\langle r,\alpha \right| \hat{S}_{x}(\mathbf{k})\hat{S}_{x}(-\mathbf{k}) \left| r,\alpha \right\rangle \,, \\ 
    \boldsymbol{J_{2}(r,\alpha)} &=& \left\langle r,\alpha \right|  \hat{S}_{x}(\mathbf{k})\hat{S}_{z}(-\mathbf{k}) \left| r,\alpha \right\rangle \,, \\ 
    \boldsymbol{J_{3}(r,\alpha)} &=& \left\langle r,\alpha \right|  \hat{S}_{z}(\mathbf{k})\hat{S}_{x}(-\mathbf{k}) \left| r,\alpha \right\rangle \,. 
\end{eqnarray}
The expectation values of the remaining three terms in the Bell operator are similarly obtained as  
\begin{eqnarray}\label{spin_three}
    E(\theta_{1},\overline{\theta}_{2}) &=& \left\langle r, \alpha \right | \hat{\vec{S}}_{\mathbf{k}}\cdot \vec{n}_{1}\otimes\hat{\vec{S}}_{-\mathbf{k}}\cdot \vec{n}^{'}_{2} \left | r, \alpha \right\rangle \nonumber \\ 
    &=& \sin{\theta_{1}}\sin{\overline{\theta}_{2}} \, \boldsymbol{J_{1}(r,\alpha)} + \cos{\theta_{1}}\cos{\overline{\theta}_{2}} \, \boldsymbol{J_{0}(r,\alpha)} + \sin{\theta_{1}}\cos{\overline{\theta}_{2}} \, \boldsymbol{J_{2}(r,\alpha)} + \cos{\theta_{1}}\sin{\overline{\theta}_{2}} \, \boldsymbol{J_{3}(r,\alpha)} \,, \\
    E(\overline{\theta}_{1},\theta_{2}) &=& \left\langle r, \alpha \right | \hat{\vec{S}}_{\mathbf{k}}\cdot \vec{n}^{'}_{1}\otimes\hat{\vec{S}}_{-\mathbf{k}}\cdot \vec{n}_{2} \left | r, \alpha \right\rangle \nonumber \\
    &=& \sin{\overline{\theta}_{1}}\sin{\theta_{2}} \, \boldsymbol{J_{1}(r,\alpha)} + \cos{\overline{\theta}_{1}}\cos{\theta_{2}} \, \boldsymbol{J_{0}(r,\alpha)} + \sin{\overline{\theta}_{1}}\cos{\theta_{2}} \, \boldsymbol{J_{2}(r,\alpha)} + \cos{\overline{\theta}_{1}}\sin{\theta_{2}} \, \boldsymbol{J_{3}(r,\alpha)} \,, \label{spin_three_1} \\
    E(\overline{\theta}_{1},\overline{\theta}_{2}) &=& \left\langle r, \alpha \right | \hat{\vec{S}}_{\mathbf{k}}\cdot \vec{n}^{'}_{1}\otimes\hat{\vec{S}}_{-\mathbf{k}}\cdot \vec{n}^{'}_{2} \left | r, \alpha \right\rangle \nonumber \\
    &=& \sin{\overline{\theta}_{1}}\sin{\overline{\theta}_{2}} \, \boldsymbol{J_{1}(r,\alpha)} + \cos{\overline{\theta}_{1}}\cos{\overline{\theta}_{2}} \, \boldsymbol{J_{0}(r,\alpha)} + \sin{\overline{\theta}_{1}}\cos{\overline{\theta}_{2}} \, \boldsymbol{J_{2}(r,\alpha)} + \cos{\overline{\theta}_{1}}\sin{\overline{\theta}_{2}} \, \boldsymbol{J_{3}(r,\alpha)} \,. \label{spin_three_2} 
\end{eqnarray}
To simplify the analysis, we exploit the freedom to choose arbitrary directions for the measurement axes defined by the unit vectors $\vec{n}_{i}$. For convenience, we fix the polar angles as $\theta_{1} = 0$, $\overline{\theta}_{1} = \pi/2$, and $\overline{\theta}_{2} = -\theta_{2}$. Substituting these values into~\ref{spin_one}, \ref{spin_three}, \ref{spin_three_1} and \ref{spin_three_2}, we obtain
\begin{eqnarray}\label{individual_exp}
    E(\theta_{1},\theta_{2}) &=& \cos{\theta_{2}} \, \boldsymbol{J_{0}(r,\alpha)} + \sin{\theta_{2}} \, \boldsymbol{J_{3}(r,\alpha)}\,, \\ 
    E(\theta_{1},\overline{\theta}_{2}) &=& \cos{\theta_{2}} \, \boldsymbol{J_{0}(r,\alpha)} - \sin{\theta_{2}} \, \boldsymbol{J_{3}(r,\alpha)} \,, \\ 
    E(\overline{\theta}_{1},\theta_{2}) &=& \sin{\theta_{2}} \, \boldsymbol{J_{1}(r,\alpha)} + \cos{\theta_{2}} \, \boldsymbol{J_{2}(r,\alpha)} \,, \\ 
    E(\overline{\theta}_{1},\overline{\theta}_{2}) &=& \cos{\theta_{2}} \, \boldsymbol{J_{2}(r,\alpha)} - \sin{\theta_{2}} \, \boldsymbol{J_{1}(r,\alpha)} \,. 
\end{eqnarray}
Now, substituting the above expressions into~\ref{Bell_operator}, we obtain the quantum expectation value of the Bell operator as 
\begin{eqnarray}\label{bell_ineq}
    \left\langle r,\alpha \right| \hat{\mathcal{B}} \left| r,\alpha \right\rangle &=& E(\theta_{1}, \theta_{2}) + E(\theta_{1}, \overline{\theta}_{2}) + E(\overline{\theta}_{1}, \theta_{2}) - E(\overline{\theta}_{1}, \overline{\theta}_{2}) \nonumber \\
    &=& 2 \, \big[\cos{\theta_{2}} \, \boldsymbol{J_{0}(r,\alpha)} + \sin{\theta_{2}} \, \boldsymbol{J_{1}(r,\alpha)} \big] \,. 
\end{eqnarray}
To test for a violation of the Bell-CHSH inequality using the squeezed coherent state $\ket{r,\alpha}$, we must determine the maxima of the Bell operator's expectation value as given in ~\ref{bell_ineq}. As the expression is a function of the angle $\theta_{2}$, the maximization is performed with respect to this variable. The optimal angle that maximizes the Bell expectation is given by
\begin{equation}
    \theta_{2,max} \, = \, \tan^{-1} \left[\frac{\boldsymbol{J_{1}(r,\alpha)}}{\boldsymbol{J_{0}(r,\alpha)}} \right] \,. 
\end{equation}
Substituting $\theta_{2,max}$ into \ref{bell_ineq}, one obtains the following form of its maxima    
\begin{eqnarray}\label{bell_inequ_max}
    \left \langle r,\alpha \right| \hat{\mathcal{B}} \left| r,\alpha \right\rangle_{max} &=& 2 \sqrt{\boldsymbol{J_{0}(r,\alpha)^{2}} + \boldsymbol{J_{1}(r,\alpha)^{2}}} \nonumber \\ 
    &=& 2 \sqrt{\left\langle r,\alpha \right|  \hat{S}_{z}(\mathbf{k})\hat{S}_{z}(-\mathbf{k}) \left| r,\alpha \right\rangle^{2} + \left\langle r,\alpha \right| \hat{S}_{x}(\mathbf{k})\hat{S}_{x}(-\mathbf{k}) \left| r,\alpha \right\rangle^{2} } \, . 
\end{eqnarray}
The remaining task is to explicitly evaluate the spin–xx and spin–zz correlations in terms of the squeezing parameters $r_k, \theta_k, \phi_k$ and the coherent state amplitudes $\alpha_{\mathbf{k}}, \alpha_{-\mathbf{k}}$. Once these expressions are obtained, one can analyze the extent of Bell-CHSH inequality violation and its dependence on the squeezed coherent state parameters.

\subsection{Spin correlations}\label{Subsection-spin-xx}

We now proceed to derive explicit expressions for the spin-spin correlations in terms of the coherent displacement and squeezing parameters. To achieve this, we begin by rewriting the pseudo-spin operators introduced in~\ref{pseudo_spin}~-~\ref{pseudo-spin_30} in terms of the eigenstates of the quadrature operator $\ket{q_{\mathbf{k}}}$ and $\ket{-q_{\mathbf{k}}}$. The resulting representations are given by
\begin{eqnarray}\label{pseudo_spin_2}
    \hat{S}_{x}(\mathbf{k}) &=& \int^{\infty}_{0} dq_{\mathbf{k}} \big(\ket{q_{\mathbf{k}}}\bra{q_{\mathbf{k}}} - \ket{-q_{\mathbf{k}}}\bra{-q_{\mathbf{k}}} \big) \,, \\ 
    \hat{S}_{y}(\mathbf{k}) &=& i\int^{\infty}_{0} dq_{\mathbf{k}} \big(\ket{q_{\mathbf{k}}}\bra{-q_{\mathbf{k}}} - \ket{-q_{\mathbf{k}}}\bra{q_{\mathbf{k}}} \big) \,, \label{spin-y} \\ 
    \hat{S}_{z}(\mathbf{k}) &=& -\int^{\infty}_{-\infty} dq_{\mathbf{k}} \ket{q_{\mathbf{k}}}\bra{-q_{\mathbf{k}}} \,. \label{spin-z}
\end{eqnarray}
First, we focus on evaluating the spin-xx correlation of the pseudo-spin operators \ref{pseudo_spin_2} evaluated with respect to squeezed coherent state $\ket{r,\alpha}$. Using the form of the pseudo-spin operator, $\hat{S}_{x}(k)$, in the quadrature basis, this correlation reduces to
\begin{eqnarray}\label{Spin xx}
    \left\langle r,\alpha \right| \hat{S}_{x}(\mathbf{k})\hat{S}_{x}(-\mathbf{k}) \left| r,\alpha \right\rangle &=& \int^{\infty}_{0} \int^{\infty}_{0} dq_{\mathbf{k}}dq_{-\mathbf{k}} \hspace{0.06cm} \big[\abs{\psi_{r}(q_{\mathbf{k}},q_{-\mathbf{k}})}^{2} + \abs{\psi_{r}(-q_{\mathbf{k}},-q_{-\mathbf{k}})}^{2} - \abs{\psi_{r}(q_{\mathbf{k}},-q_{-\mathbf{k}})}^{2} \nonumber \\ 
    && \hspace{0.7cm} - \hspace{0.1cm} \abs{\psi_{r}(-q_{\mathbf{k}},q_{-\mathbf{k}})}^{2} \big] \,.   
\end{eqnarray}
In the above expression, $\psi_{r}(q_{\mathbf{k}},q_{-\mathbf{k}})$ represents the wave function of squeezed coherent state in the quadrature basis $(q_{\mathbf{k}}, q_{-\mathbf{k}})$, which is given in \ref{Wave function}.

Finally, using the wave function~\ref{Wave function} we obtain the spin-xx correlation, \ref{Spin xx} of squeezed coherent state as 
\begin{eqnarray}\label{Spin-xx Correlation}
    \left\langle r,\alpha \right| \hat{S}_{x}(\mathbf{k})\hat{S}_{x}(-\mathbf{k}) \left| r,\alpha \right\rangle &=& \frac{4\pi\abs{M_{0}}^{2}}{\sqrt{g_{1} g_{2}}} \left\{\OwenT{\left[\hspace{0.1cm} \frac{\sqrt{2 g_{1} g_{2}} \, (y_{0}-x_{0})}{\sqrt{g_{1}+g_{2}}} \boldsymbol{;} \frac{\left(g_{1}x_{0}+g_{2}y_{0} \right)}{\sqrt{g_{1}g_{2}} \, (y_{0}-x_{0})} \right]} + \frac{1}{8} \Sign{\left[\sqrt{g_{2}} \left(\frac{y_{0}}{x_{0}}+1 \right) \right]} \right. \nonumber \\ 
    && \hspace{1.1cm} - \hspace{0.1cm} \OwenT{\left[\hspace{0.1cm} \frac{\sqrt{2 g_{1}g_{2}} \, (y_{0}+x_{0})}{\sqrt{g_{1}+g_{2}}} \boldsymbol{;} \frac{\left(g_{1}x_{0}-g_{2}y_{0} \right)}{\sqrt{g_{1}g_{2}} \, (y_{0}+x_{0})} \right]} - \frac{1}{8} \Sign{\left[\sqrt{g_{2}} \left(\frac{y_{0}}{x_{0}}-1 \right) \right]} \nonumber \\ 
    && \left. \hspace{1.2cm} + \hspace{0.1cm} \frac{1}{8} \Sign{\left[\sqrt{g_{1}} \left(\frac{x_{0}}{y_{0}}-1 \right) \right]} - \frac{1}{8} \Sign{\left[\sqrt{g_{1}} \left(\frac{x_{0}}{y_{0}}+1 \right) \right]} \right\} \,. 
\end{eqnarray}
For a step-by-step derivation of the above result, see the Appendix~\ref{Spin-xx_calculation}. 
The quantities $g_{1}, g_{2}, x_{0}, y_{0}$ and $M_{0}$ appearing in \ref{Spin-xx Correlation}, are defined as follows:
\begin{eqnarray}
    g_{1} \hspace{-0.2cm} &=& \frac{1}{2} \left[2\Re\big(A_{0} \big)-\Re\big(B_{0} \big) \right] \hspace{0.2cm} = \hspace{0.2cm} \frac{\sech^{2}{r_{k}}}{2} \bigg(\frac{1 + \tanh^{2}{r_{k}} - 2\cos{2\phi_{k}} \, \tanh{r_{k}}}{1 + \tanh^{4}{r_{k}} - 2\cos{4\phi_{k}} \, \tanh^{2}{r_{k}}} \bigg) \,, \label{def-g_1} \\ 
    g_{2} \hspace{-0.2cm} &=& \frac{1}{2} \left[2\Re\big(A_{0} \big)+\Re\big(B_{0} \big) \right] \hspace{0.2cm} = \hspace{0.2cm} \frac{\sech^{2}{r_{k}}}{2} \bigg(\frac{1 + \tanh^{2}{r_{k}} + 2\cos{2\phi_{k}} \, \tanh{r_{k}}}{1 + \tanh^{4}{r_{k}} - 2\cos{4\phi_{k}} \, \tanh^{2}{r_{k}}} \bigg) \,, \label{def-g_2} \\ 
    x_{0} \hspace{-0.2cm} &=& \frac{1}{g_{1}} \left[\Re\big(A_{0}C_{0} \big) + \Re\big(A_{0}D_{0} \big) \right]  \nonumber \\ 
    \hspace{-0.2cm} &=& \frac{4\cosh^{2}{r_{k}}}{\big(1 + \tanh^{2}{r_{k}} - 2\cos{2\phi_{k}} \, \tanh{r_{k}} \big)} \left\{\frac{1}{2}\Im\big(\beta_{\mathbf{k}}+\gamma_{\mathbf{k}} \big) \sin{4\phi_{k}} \, \tanh^{2}{r_{k}} + \right. \nonumber \\ 
    && \left. + \hspace{0.2cm} \frac{1}{2}\Re\big(\beta_{\mathbf{k}}+\gamma_{\mathbf{k}} \big) \left[\Re\big(A_{0} \big)\sech^{2}{r_{k}} - 2\Im\big(A_{0} \big)\sin{2\phi_{k}} \, \tanh{r_{k}} \right] \big(1 + \tanh^{2}{r_{k}} - 2\cos{2\phi_{k}} \, \tanh{r_{k}} \big) \right\} \,, \label{def-x_0} \\ 
    y_{0} \hspace{-0.2cm} &=& \frac{1}{g_{2}} \left[\Re\big(A_{0}C_{0} \big) - \Re\big(A_{0}D_{0} \big) \right] \nonumber \\ 
    \hspace{-0.2cm} &=& \frac{4\cosh^{2}{r_{k}}}{\big(1 + \tanh^{2}{r_{k}} + 2\cos{2\phi_{k}} \, \tanh{r_{k}} \big)} \left\{\frac{1}{2} \Im\big(\beta_{\mathbf{k}}-\gamma_{\mathbf{k}} \big) \sin{4\phi_{k}}\tanh^{2}{r_{k}} + \right. \nonumber \\ 
    && \left. + \hspace{0.2cm} \frac{1}{2} \Re\big(\beta_{\mathbf{k}}-\gamma_{\mathbf{k}} \big) \left[\Re\big(A_{0} \big)\sech^{2}{r_{k}} + 2\Im\big(A_{0} \big)\sin{2\phi_{k}} \, \tanh{r_{k}} \right] \big(1 + \tanh^{2}{r_{k}} + 2\cos{2\phi_{k}} \, \tanh{r_{k}} \big) \right\} \,, \label{def-y_0} \\ 
    \abs{M_{0}}^{2} \hspace{-0.1cm} &=& \frac{\abs{N_{0}}^{2}}{2} \exp\left\{-2 \Re\left[A_{0}\big(C_{0}^{2}+D_{0}^{2} \big) \right] + g_{1} x^{2}_{0} + g_{2} \, y^{2}_{0} \, \right\}  \,. \label{def-M_0} \hspace{0.6 cm}
\end{eqnarray}

It is interesting to note that $\left\langle r,\alpha \right| \hat{S}_{x}(\mathbf{k})\hat{S}_{x}(-\mathbf{k}) \left| r,\alpha \right\rangle$ can be obtained analytically in terms of two special functions: Owen’s T-function and the Signum function. These functions are defined as
\begin{eqnarray}
    \OwenT{(x, h)} &=& \frac{1}{2\pi} \int^{h}_{0} dt \, \frac{e^{-x^{2}(1 + t^{2})/2}}{(1 + t^{2})} \, , \\ 
    \Sign{(x)} &=& 1 \hspace{0.8 cm} \text{for} \hspace{0.3 cm} x > 0 \, , \nonumber \\ 
             &=& 0 \hspace{0.8 cm} \text{for} \hspace{0.3 cm} x = 0 \, , \nonumber \\ 
             &=& -1 \hspace{0.6 cm} \text{for} \hspace{0.3 cm} x < 0 \, . 
\end{eqnarray}
The behavior of Owen’sT-function with respect to its first argument closely resembles that of a Gaussian distribution centered at the origin. For different values of its second argument, the height of the peak changes accordingly, while its location and overall shape remain unaffected. 

\color{black}

According to the GKMR prescription (see ~\ref{spin-z}), the spin–zz correlation is always found to be equal to unity:
\begin{equation}
\left\langle r,\alpha \right|  \hat{S}_{z}(\mathbf{k})\hat{S}_{z}(-\mathbf{k}) \left| r,\alpha \right\rangle \hspace{0.3 cm} = \hspace{0.3 cm} 1 \, .    
\end{equation}
Finally, the maximum quantum expectation value of the Bell operator, as defined in \ref{bell_inequ_max}, is given by 
\begin{eqnarray}\label{bell_ineu_SC}
    \left\langle r,\alpha \right| \hat{\mathcal{B}} \left| r,\alpha \right\rangle_{max} &=& 2 \, \sqrt{1 + \left\langle r,\alpha \right| \hat{S}_{x}(\mathbf{k})\hat{S}_{x}(-\mathbf{k}) \left| r,\alpha \right\rangle^{2}} \, , 
\end{eqnarray}
where spin-xx correlation of squeezed coherent state is given by \ref{Spin-xx Correlation}. At this stage, from \ref{bell_ineu_SC}, it is possible to comment on the bounds of the quantum-averaged Bell operator. According to the GKMR prescription (see~ \ref{pseudo_spin_2}), the pseudo-spin operators are constructed such that their eigenvalues are strictly $\pm 1$. As a result, the expectation values of all spin–spin correlations are confined within the interval $[-1, 1]$. In particular, the spin–zz correlation is always equal to 1, while the spin–xx correlation varies between 0 and 1 depending on the state parameters. The Bell operator reaches its maximum value of $2\sqrt{2}$ when the spin–xx correlation attains its upper bound of 1. Conversely, when the spin–xx correlation vanishes, the Bell operator takes its minimum value of 2. From this reasoning, one can immediately conclude that the quantum-averaged Bell operator is bounded as: $2 \le \langle \hat{\mathcal{B}} \rangle \le 2\sqrt{2}
$, clearly demonstrating the window for Bell inequality violation in the context of the squeezed coherent state.

As discussed earlier, the central goal of this work is to investigate the role of coherent state parameters $\alpha_{\mathbf{k}}$ and $\alpha_{-\mathbf{k}}$ in the violation of the Bell inequality~\ref{bell_ineu_SC}. These parameters, in conjunction with the squeezing parameters, appear in the spin–xx correlation expression~\ref{Spin-xx Correlation} through the arguments of special functions such as Owen’s $T$ function and the Signum function. Specifically, the quantities $x_0$ and $y_0$ encapsulate the explicit dependence on the coherent state parameters. As a result, we find that the Bell inequality becomes sensitive to the values of $\alpha_{\mathbf{k}}$ and $\alpha_{-\mathbf{k}}$ when the spin–spin correlations are evaluated with respect to the squeezed coherent state $\ket{r, \alpha}$.

\subsection{Bell Inequality violation in de Sitter inflation}

The inflationary phase of the universe is often approximated by a de Sitter expansion, primarily because it allows for analytical solutions of perturbations. Although current observations indicate slight deviations from exact de Sitter behavior, the de Sitter approximation remains a useful and widely adopted model in early universe cosmology. In the case of de Sitter inflation, the time-dependent squeezing parameter $r_{k}, \theta_{k}$ and $\phi_{k}$ admit closed-form expressions, as derived in\cite{Martin_2016}, 
\begin{eqnarray}\label{sol_de_sitter}
    r_{k}(\eta) &=& - \sinh^{-1}{\left(\frac{1}{2k\eta} \right)} \,, \\ 
    \phi_{k}(\eta) &=& \frac{\pi}{4} - \frac{1}{2} \tan^{-1}{\left(\frac{1}{2k\eta} \right)} \,, \label{squeezing} \\ 
    \theta_{k}(\eta) &=& k\eta + \tan^{-1}{\left(\frac{1}{2k\eta} \right)} \,\label{rotation} . 
\end{eqnarray}
It is evident from the expressions above that the squeezing parameters are monotonic functions of conformal time $\eta$. This monotonicity allows one to express $\theta_k$ and $\phi_k$ as functions of the squeezing parameter $r_k$, thereby enabling the use of $r_k$ itself as an effective time variable for studying the evolution of the expectation value of the Bell operator. Furthermore, the squeezing parameter $r_{k}$ starts from zero when the modes are well inside the Hubble radius and grows continuously, becoming larger than unity when the corresponding mode exits the Hubble radius, i.e., when $k < aH$.

\begin{figure}[h!]
    \centering
        \includegraphics[width=0.49\linewidth]{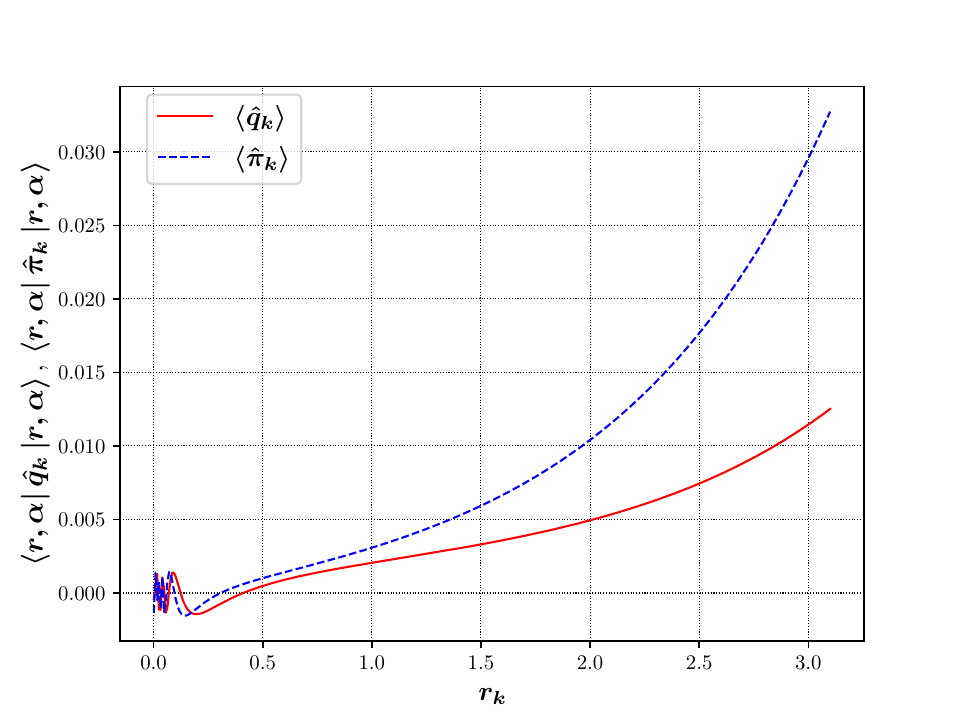}    \includegraphics[width=0.49\linewidth]{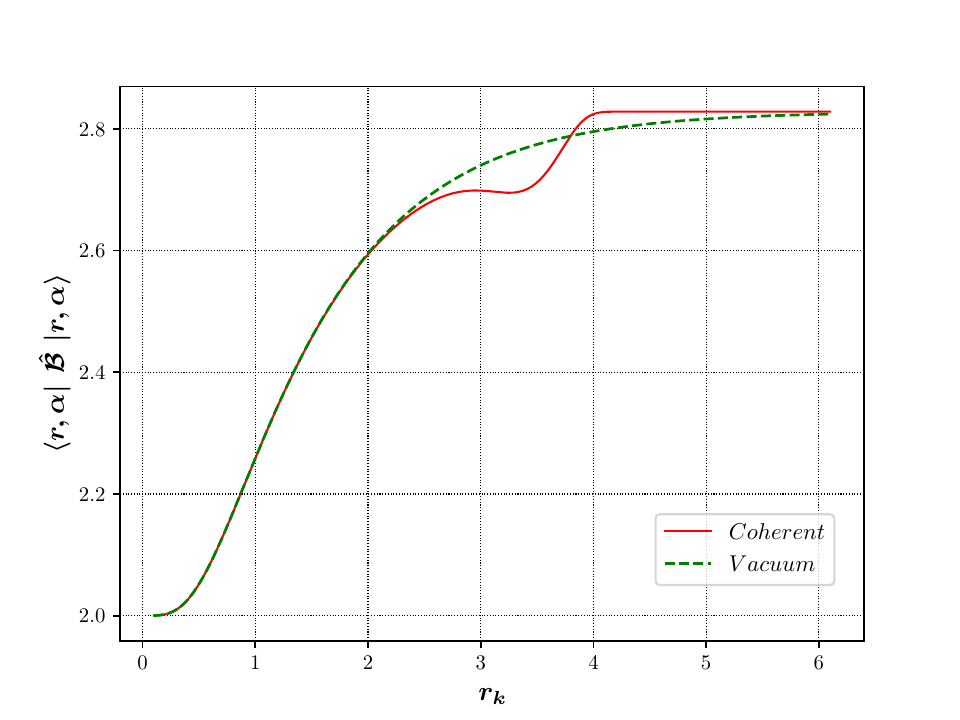} \\
        \includegraphics[width=0.49\linewidth]{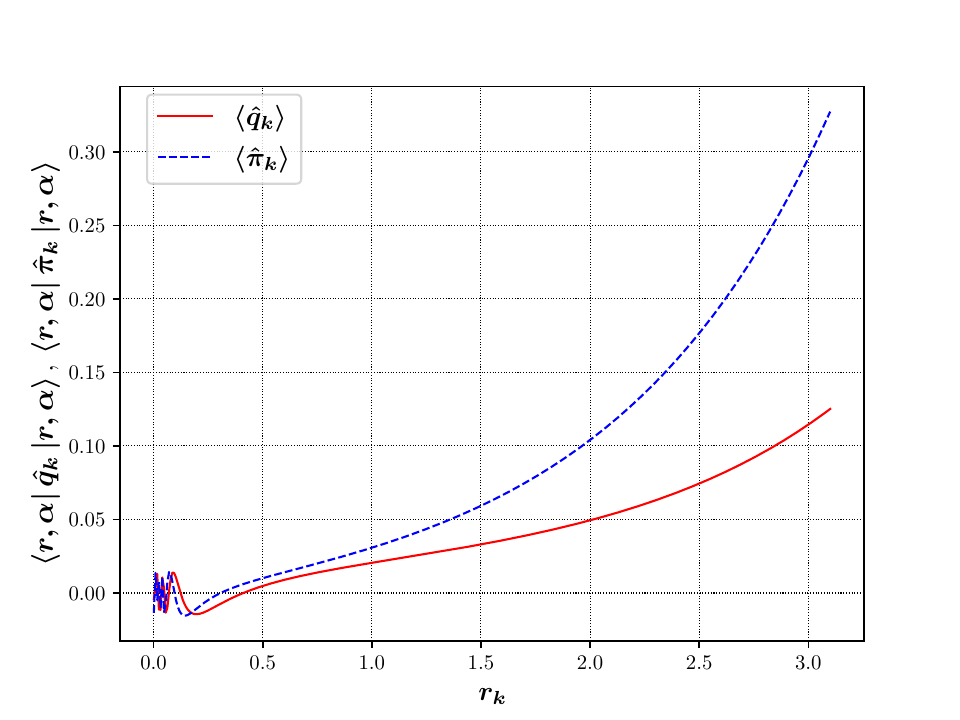} 
       \includegraphics[width=0.49\linewidth]{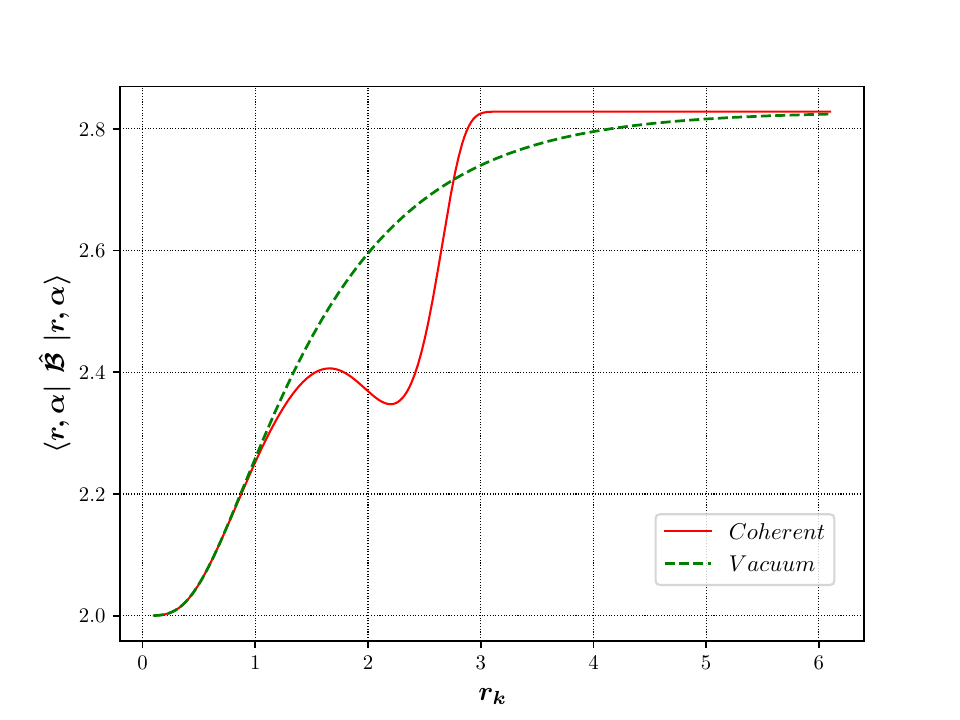}

        \includegraphics[width=0.49\linewidth]{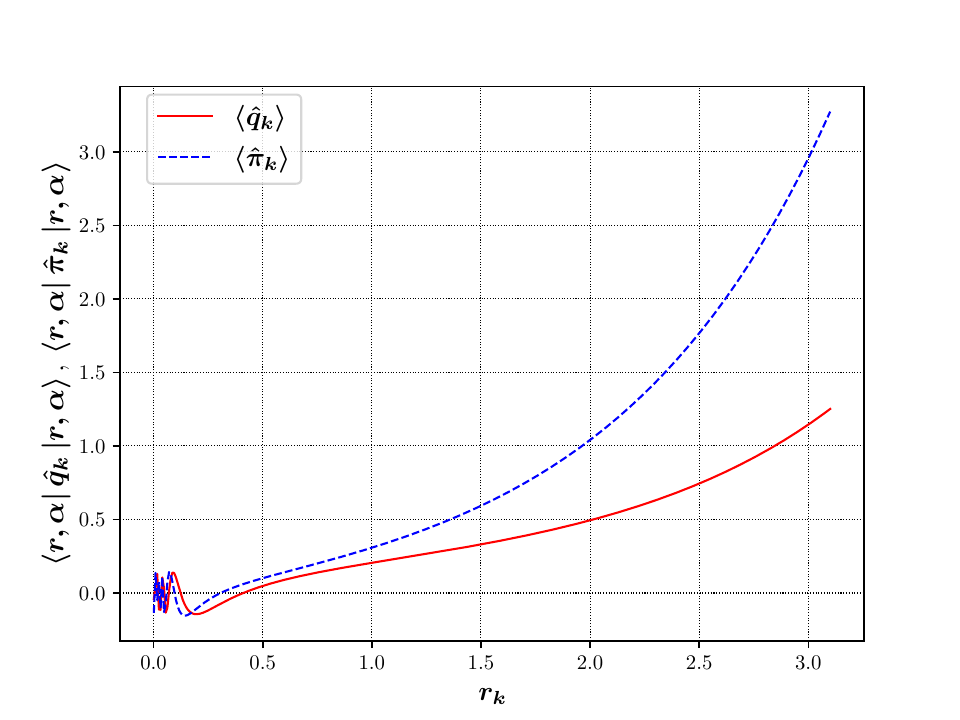} 
        \includegraphics[width=0.49\linewidth]{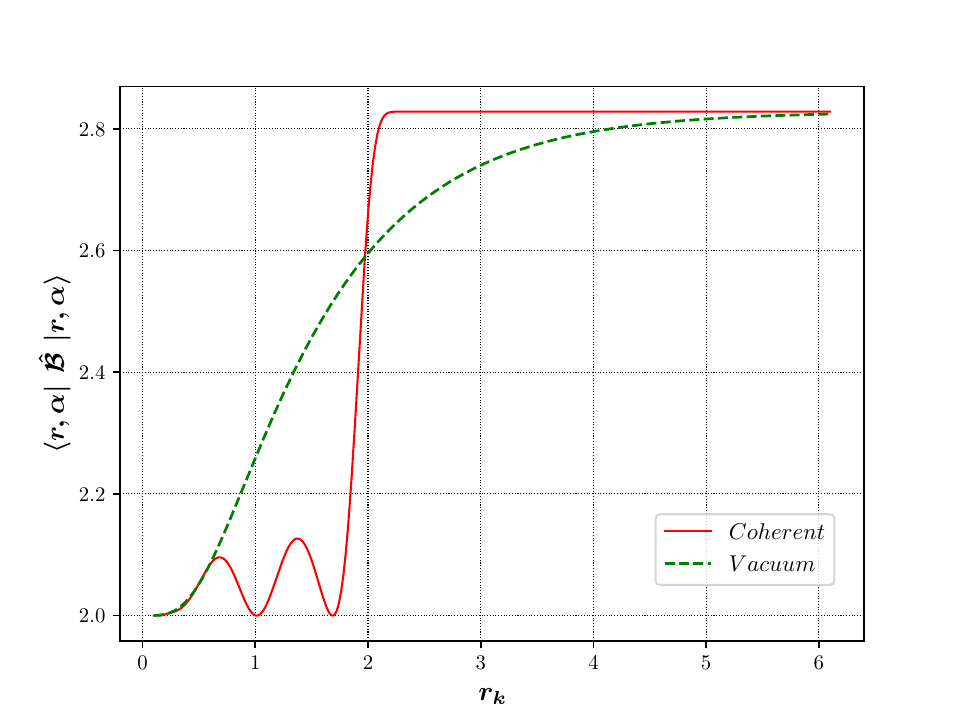}
        \captionsetup{width=0.9\linewidth}
        \caption{\small{One-point correlation of $\hat{q}_{\mathbf{k}}, \hat{\pi}_{\mathbf{k}}$ (on the left column) and the expectation value of Bell operator (on the right column) as a function of $r_{k}$ while coherent state parameters are fixed as, $\alpha_{\mathbf{k}} = 0.001 + i\,0.001$ and $\alpha_{-\mathbf{k}} = 0.002 + i\,0.002$ (on the first row), $\alpha_{\mathbf{k}} = 0.01 + i\,0.01$ and $\alpha_{-\mathbf{k}} = 0.02 + i\,0.02$ (on the second row), $\alpha_{\mathbf{k}} = 0.1 + i\,0.1$ and $\alpha_{-\mathbf{k}} = 0.2 + i\,0.2$ (on the third row). The green curves represent the results of squeezed vacuum state i.e. $\alpha_{\mathbf{k}} = \alpha_{-\mathbf{k}} = 0$. The coherent-state parameters are chosen such that increasing values of $\alpha_{\mathbf{k}}$ correspond to increasing magnitudes of the one-point correlation function.} }
        \label{fig:q-p-avg}
\end{figure}

The left-hand side of \ref{fig:q-p-avg} shows the evolution of the expectation values of $\hat{q}_{\mathbf{k}}$ and $\hat{\pi}_{\mathbf{k}}$ as functions of the squeezing parameter $r_{k}$. The plots indicate that the single-point expectation values increase as inflation progresses. Moreover, the expectation values grow with increasing values of the coherent state parameters $\alpha_{\mathbf{k}}$ and $\alpha_{-\mathbf{k}}$. On the right-hand side of \ref{fig:q-p-avg}, we plot the expectation value of the Bell operator evaluated with respect to the squeezed coherent state, $\left\langle r,\alpha \right| \hat{\mathcal{B}} \left| r,\alpha \right\rangle$, as a function of the squeezing parameter $r_k$, keeping the coherent state parameters $\alpha_{\mathbf{k}}$ and $\alpha_{-\mathbf{k}}$ fixed. The green and red curves represent the behavior of the quantum-averaged Bell operator for the squeezed vacuum and squeezed coherent states, respectively. For smaller values of $r_k$, the two curves exhibit distinct behavior, as shown in \ref{fig:q-p-avg}. However, for large values of $r_k$, corresponding to the super-Hubble regime, the Bell operator expectation value for the squeezed coherent state $\ket{r,\alpha}$ asymptotically approaches that of the squeezed vacuum state $\ket{r,0}$. This indicates that, although the choice of initial state influences the behavior of $\left\langle r,\alpha \right| \hat{\mathcal{B}} \left| r,\alpha \right\rangle$ during the early stages of evolution, the distinction fades in the super-Hubble limit.

The saturation behavior of $\left\langle r,\alpha \right| \hat{\mathcal{B}} \left| r,\alpha \right\rangle$ can be understood analytically from the expression in \ref{Spin-xx Correlation}. In the super-Hubble limit ($k\eta \ll 0$), where the squeezing parameter $r_k$ becomes very large, the first argument of both Owen’s~T functions appearing in the spin-xx correlation tends toward infinity. Meanwhile, the second argument of these Owen’s~T functions remain finite and depends on the coherent state parameters $\alpha_{\mathbf{k}}$ and $\alpha_{-\mathbf{k}}$. Consequently, both Owen's~T functions vanish in this limit due to the divergent first argument. On the other hand, the Signum functions in \ref{Spin-xx Correlation} contribute non-trivially as $r_k \to \infty$. A careful analysis shows that the combined contribution from all the four Signum functions amounts to 2. Summing up all these contributions, one finds that the expectation value of the Bell operator asymptotically reaches the Tsirelson bound of Bell inequality, i.e., $2\sqrt{2}$, in the super-Hubble limit \cite{Cirelson:1980ry}. 

Another notable inference from these plots is that the saturation to a constant value occurs more rapidly for $\alpha_{\mathbf{k}} \neq 0$ compared to the vacuum case. Moreover, this saturation becomes faster as the magnitude of $\alpha_{\mathbf{k}}$ increases. This suggests that an increase in the one-point correlation function enhances the rate at which the Bell inequality reaches its maximal violation bound.

\color{black}
\section{Conclusion}\label{sec:8}

It is often assumed that primordial perturbations originate as quantum fluctuations during inflation. These fluctuations eventually become classical and leave imprints in the anisotropies of the Cosmic Microwave Background (CMB) temperature. Several efforts have been made in the literature to investigate the quantum nature of these perturbations. One powerful way to characterize quantum correlations is through the violation of Bell inequalities. Most of these studies assume that primordial fluctuations evolve from the Bunch-Davies vacuum state, though this is not the only possibility. In this work, we consider a coherent state as the initial state of primordial perturbations and examine the associated Bell inequalities to explore their quantum properties.

If the evolution starts from the Bunch-Davies vacuum, the quantum state evolves into a squeezed vacuum state on super-Hubble scales. Similarly, if the initial state is a coherent state, the evolution leads to a squeezed coherent state. In this work, we derive a general analytical expression for the expectation value of the Bell operator—constructed from the pseudo-spin operators introduced in~\ref{pseudo_spin_2}—with respect to the squeezed coherent state. We then analyze the evolution of expectation value of the Bell operator value as a function of the squeezing parameters in the context of de Sitter inflation and compare it with the vacuum case.

Our analysis shows that, although the expectation value of the Bell operator for cosmological perturbations initially deviates from that of the vacuum case, it asymptotically converges to the same value. This indicates that, toward the end of inflation, the Bell operator expectation value saturates to a constant for both squeezed vacuum and squeezed coherent states. However, this saturation occurs more rapidly for the case of squeezed coherent state, and the rate of convergence increases with the magnitude of $\alpha_{\mathbf{k}}$. This suggests that a larger one-point correlation function of the cosmological perturbations accelerates the approach to the maximal violation bound of the Bell inequality in its CHSH form. In other words, when the perturbations are prepared in a coherent state, the system exhibits its strongest quantum correlations—and hence its maximal quantumness—earlier in its evolution compared to the vacuum case.

Before concluding, it is important to mention that, even though formulating a Bell-inequality test for primordial perturbations is conceptually straightforward, implementing such a test using observational data is extremely challenging. The main obstacle is that current CMB measurements provide access only to the two-point correlation functions of the Mukhanov–Sasaki variable, but not to those of its conjugate momentum. The expectation values of the pseudo-spin operators and the associated Bell operator depend on both the field and momentum correlators. Since momentum correlations are not accessible with present CMB data, a direct test of Bell-inequality violations using CMB observations alone remains infeasible.

While our analysis has primarily focused on cosmological perturbations in Fourier space, observational data is inherently obtained in real space. Therefore, it is important to investigate quantum mechanical tests, such as Bell inequality violations, directly in real space. We plan to extend our current work in this direction. Moreover, in the alternative approach proposed in Refs.~\cite{Dale:2025nhc, Dale:2023fnp}, the Bell operator is constructed directly from the temperature fluctuations in the CMB. Recently, using Planck data, these works report a robust violation of a cosmic CHSH inequality. The key difference between our work and this CMB-based proposal is that we focus on the primordial perturbations and their quantum evolution during inflation, whereas the latter approach deals with the late-time temperature anisotropies observed in the CMB.

Additionally, our study has focused on a specific pair of operators, namely $q_{\mathbf{k}}$ and $q_{-\mathbf{k}}$. However, various other canonical combinations of these variables can be considered, and it is well known that entanglement entropy depends on the choice of canonical variables. As a future direction, we aim to explore how the violation of Bell inequalities is affected by canonical transformations of the variables associated with cosmological perturbations.


\section*{Acknowledgments}

The authors wish to thank Sumanta Chakraborty for interesting discussions. RNR wishes to acknowledge the support from Anusandhan National Research Foundation (ANRF), Science and Engineering Research Board (SERB), Department of Science and Technology (DST), Government of India, through National Post-Doctoral Fellowship PDF/2023/001226. \\

\appendix
\noindent\textbf{\Large Appendix}

\section{Wave function of squeezed coherent state}\label{appendix:B}

In this section, we derive the wave function of the squeezed coherent state in the quadrature basis $\ket{q_{\mathbf{k}}, q_{-\mathbf{k}}}$. The squeezed coherent state is defined as the sequential action of the squeezing operator $\hat{S}_{\mathbf{k}}(r_{k}, \phi_{k})$, rotation operator $\hat{R}_{\mathbf{k}}(\theta_{k})$, and displacement operator $\hat{D}(\alpha_{\mathbf{k}}, \alpha_{-\mathbf{k}})$ on the two-mode vacuum state $\ket{0_{\mathbf{k}}, 0_{-\mathbf{k}}}$. Let us begin by evaluating the action of the displacement operator,
\begin{eqnarray}
    \hat{D}(\alpha_{\mathbf{k}}) &=& \exp\big(\alpha_{\mathbf{k}} \, \hat{a}^{\dagger}_{\mathbf{k}} - \alpha^{*}_{\mathbf{k}} \, \hat{a}_{\mathbf{k}} \big) \hspace{0.3 cm} = \hspace{0.3 cm} \exp\left\{2i\left[\Im(\alpha_{\mathbf{k}}) \, \hat{q}_{\mathbf{k}} - \Re(\alpha_{\mathbf{k}}) \, \hat{\pi}_{\mathbf{k}} \right] \right\} , 
\end{eqnarray}
on the complex conjugate of the quadrature basis $\bra{q_{\mathbf{k}}, q_{-\mathbf{k}}}$ as
\begin{eqnarray}
    \bra{q_{\mathbf{k}}} \hat{D}(\alpha_{\mathbf{k}}) 
    &=& e^{2i\Im(\alpha_{\mathbf{k}} )q_{\mathbf{k}}} \hspace{0.1cm} e^{-2i\hbar\Re(\alpha_{\mathbf{k}} )\Im(\alpha_{\mathbf{k}} )} \bra{q_{\mathbf{k}}-2\Re\big(\alpha_{\mathbf{k}} \big) } \hspace{0.06cm} .
\end{eqnarray}
Similarly, the displacement operator acting on the other mode gives: 
\begin{equation}
    \bra{q_{-\mathbf{k}}} \hat{D}(\alpha_{-\mathbf{k}}) \hspace{0.2 cm} = \hspace{0.2 cm} e^{2i\Im(\alpha_{-\mathbf{k}} )q_{-\mathbf{k}}} \hspace{0.1cm} e^{-2i\hbar\Re(\alpha_{-\mathbf{k}} )\Im(\alpha_{-\mathbf{k}} )} \bra{q_{-\mathbf{k}} - 2\Re\big(\alpha_{-\mathbf{k}} \big) } \hspace{0.06cm} .
\end{equation}
To simplify the computation, we perform a change of variables from $(q_{\mathbf{k}}, q_{-\mathbf{k}})$ to $(y_{\mathbf{k}}, y_{-\mathbf{k}})$, defined by
\begin{equation}
    y_{\mathbf{k}} \, := \, q_{\mathbf{k}}-2\Re\big(\alpha_{\mathbf{k}} \big) \, , \quad y_{-\mathbf{k}} \, := \, q_{-\mathbf{k}}-2\Re\big(\alpha_{-\mathbf{k}} \big).
\end{equation}
Next, we compute the following matrix element  $\langle y_{\mathbf{k}},y_{-\mathbf{k}} | \hat{S}_{\mathbf{k}}(r_{k}, \phi_{k}) \hat{R}_{\mathbf{k}}(\theta_{k}) | 0_{\mathbf{k}}, 0_{-\mathbf{k}}\rangle$ :
\begin{eqnarray}\label{inter_wave_function}
    \big{\langle} y_{\mathbf{k}}, y_{-\mathbf{k}} \big{|} \hat{S}_{\mathbf{k}}(r_{k}, \phi_{k}) \hat{R}_{\mathbf{k}}(\theta_{k}) \big{|} 0_{\mathbf{k}}, 0_{-\mathbf{k}} \big{\rangle} &=& \big{\langle} y_{\mathbf{k}}, y_{-\mathbf{k}} \big{|} \, \exp\left(e^{-2i\phi_{k}}\tanh{r_{k}} \, \hat{a}^{\dagger}_{\mathbf{k}}\hat{a}^{\dagger}_{-\mathbf{k}} \right) \, \exp\left[-\ln{(\cosh{r_{k}})} \big(\hat{a}^{\dagger}_{\mathbf{k}}\hat{a}_{\mathbf{k}} + \hat{a}_{-\mathbf{k}}\hat{a}^{\dagger}_{-\mathbf{k}} \big)\right] \nonumber \\ 
    && \cross \exp\left(-e^{2i\phi_{k}}\tanh{r_{k}} \, \hat{a}_{\mathbf{k}}\hat{a}_{-\mathbf{k}} \right) \, \big{|} 0_{\mathbf{k}}, 0_{-\mathbf{k}} \big{\rangle} \nonumber \\ 
    &=& \sech{r_{k}} \sum_{n} \frac{1}{n!} \big(e^{-2i\phi_{k}}\tanh{r_{k}} \big)^{n} \hspace{0.1cm} \big{\langle} y_{\mathbf{k}}, y_{-\mathbf{k}} \big{|} \, \big(\hat{a}^{\dagger}_{\mathbf{k}}\hat{a}^{\dagger}_{-\mathbf{k}} \big)^{n} \, \big{|} 0_{\mathbf{k}}, 0_{-\mathbf{k}} \big{\rangle} \nonumber \\ 
    &=& \frac{\sech{r_{k}}}{\sqrt{\pi}} \hspace{0.1cm} \exp\left[-\frac{1}{2} \big(y^{2}_{\mathbf{k}}+y^{2}_{-\mathbf{k}} \big) \right] \sum_{n} \frac{\big(e^{-2i\phi_{k}}\tanh{r_{k}}/2 \big)^{n}}{n!} \hspace{0.1cm} H_{n}(y_{\mathbf{k}}) H_{n}(y_{-\mathbf{k}}) \nonumber \\
    &=& \frac{\sech{r_{k}}}{\sqrt{\pi(1-4w^{2})}} \hspace{0.1cm} \exp\left[-\bigg(\frac{1+4w^{2}}{1-4w^{2}} \bigg) \big(y^{2}_{\mathbf{k}} + y^{2}_{-\mathbf{k}}\big) + \bigg(\frac{4w}{1-4w^{2}} \bigg) y_{\mathbf{k}}y_{-\mathbf{k}} \right] \hspace{0.06cm} . \hspace{-1.6cm}
\end{eqnarray}
The parameter~$w$ is defined as, $w := \frac{1}{2} e^{-2i\phi_{k}} \tanh{r_{k}}$. Apart from this, we used the following identity (see Appendix of \cite{Martin_2016}): 
\begin{equation}
    \sum^{\infty}_{n=0} \frac{(w)^{n}}{n!} H_{n}(y_{\mathbf{k}})H_{n}(y_{-\mathbf{k}}) \, = \, \frac{1}{\sqrt{1-4w^{2}}} \exp\left[\frac{4w^{2}}{(4w^{2}-1)}\big(y^{2}_{\mathbf{k}}+y^{2}_{-\mathbf{k}} \big) - \frac{4w}{(4w^{2}-1)} \hspace{0.1cm} y_{\mathbf{k}}y_{-\mathbf{k}}\right] \hspace{0.06cm} . 
\end{equation}
Now we compute the wave function of squeezed coherent state in quadrature basis, 
\begin{eqnarray}\label{wave_function_SC}
    \psi_{r}(q_{\mathbf{k}},q_{-\mathbf{k}}) &=& \big{\langle} q_{\mathbf{k}},q_{-\mathbf{k}} \, \big{|} \, r,\alpha \big{\rangle} \nonumber \\
    &=& \big{\langle} q_{\mathbf{k}},q_{-\mathbf{k}} \big{|} \hspace{0.1cm} \hat{S}_{\mathbf{k}}(r_{k},\phi_{k}) \hat{R}_{\mathbf{k}}(\theta_{k}) \hspace{0.1cm} \hat{D}(\alpha_{\mathbf{k}})\hat{D}(\alpha_{-\mathbf{k}}) \hspace{0.1cm} \big{|} 0_{\mathbf{k}}, 0_{-\mathbf{k}} \big{\rangle} \nonumber \\
    &=& \big{\langle} q_{\mathbf{k}},q_{-\mathbf{k}} \big{|} \hspace{0.1cm} \hat{D}(\beta_{\mathbf{k}})\hat{D}(\gamma_{\mathbf{k}}) \hspace{0.1cm} \hat{S}_{\mathbf{k}}(r_{k},\phi_{k}) \hat{R}_{\mathbf{k}}(\theta_{k}) \hspace{0.1cm} \big{|} 0_{\mathbf{k}}, 0_{-\mathbf{k}} \big{\rangle} \nonumber \\
    &=& \frac{\sech{r_{k}}}{\sqrt{\pi(1-4w^{2})}} \hspace{0.1cm} \exp\left[-A_{0}\big(q_{\mathbf{k}}-C_{0} \big)^{2} - A_{0}\big(q_{-\mathbf{k}}-D_{0} \big)^{2} + B_{0} \hspace{0.1cm} q_{\mathbf{k}} \, q_{-\mathbf{k}} + \Sigma_{0} \right] \hspace{0.06cm} ,
\end{eqnarray}
where the coefficients are defined as 
\begin{eqnarray}
    A_{0} &=& \frac{1}{2} \bigg(\frac{1+e^{-4i\phi_{k}}\tanh^{2}{r_{k}}}{1-e^{-4i\phi_{k}}\tanh^{2}{r_{k}}} \bigg) \hspace{0.06cm} , \\
    B_{0} &=& \bigg(\frac{2 \hspace{0.1cm} e^{-2i\phi_{k}}\tanh{r_{k}}}{1-e^{-4i\phi_{k}}\tanh^{2}{r_{k}}} \bigg) \hspace{0.06cm} , \\
    C_{0} &=& \frac{1}{\big(1 - e^{-4i\phi_{k}}\tanh^{2}{r_{k}} \big)} \left[\big(\beta_{\mathbf{k}} + e^{-4i\phi_{k}}\tanh^{2}{r_{k}} \hspace{0.06cm} \beta^{*}_{\mathbf{k}} \big) - 2 e^{-2i\phi_{k}}\tanh{r_{k}} \Re\big(\gamma_{\mathbf{k}} \big) \right] \hspace{0.06cm} , \\ 
    D_{0} &=& \frac{1}{\big(1 - e^{-4i\phi_{k}}\tanh^{2}{r_{k}} \big)} \left[\big(\gamma_{\mathbf{k}} + e^{-4i\phi_{k}}\tanh^{2}{r_{k}} \hspace{0.06cm} \gamma^{*}_{\mathbf{k}} \big) - 2e^{-2i\phi_{k}}\tanh{r_{k}} \Re\big(\beta_{\mathbf{k}} \big) \right] \hspace{0.06cm} , \\ 
    \Sigma_{0} &=& \hspace{-0.2 cm} \left\{A_{0}\left[C^{2}_{0}+D^{2}_{0}- 4\Re\big(\beta_{\mathbf{k}} \big)^{2} - 4\Re\big(\gamma_{\mathbf{k}} \big)^{2} \right] + 4B_{0} \Re\big(\beta_{\mathbf{k}} \big)\Re\big(\gamma_{\mathbf{k}} \big) - 2i \left[\Re\big(\beta_{\mathbf{k}} \big)\Im\big(\beta_{\mathbf{k}} \big) + \Re\big(\gamma_{\mathbf{k}} \big) \Im\big(\gamma_{\mathbf{k}} \big) \right] \right\} , \hspace{0.65 cm} \\ 
    \beta_{\mathbf{k}} &=& \big(e^{-i\theta_{k}}\cosh{r_{k}} \hspace{0.1cm} \alpha_{\mathbf{k}} - \hspace{0.1cm} e^{i(\theta_{k}+2\phi_{k})}\sinh{r_{k}} \hspace{0.1cm} \alpha^{*}_{-\mathbf{k}} \big) \hspace{0.06cm} , \\ 
    \gamma_{\mathbf{k}} &=& \big(e^{-i\theta_{k}}\cosh{r_{k}} \hspace{0.1cm} \alpha_{-\mathbf{k}} - \hspace{0.1cm} e^{i(\theta_{k}+2\phi_{k})}\sinh{r_{k}} \hspace{0.1cm} \alpha^{*}_{\mathbf{k}} \big) \hspace{0.06cm} . \hspace{1.8cm} 
\end{eqnarray}
Note that, there exist a minor discrepancy between the expressions of squeezed coherent state wave function as presented in the \ref{Wave function} and \ref{wave_function_SC}. It basically arises due to the fact that later \ref{wave_function_SC} has not been normalized to unity, whereas the former \ref{Wave function} has done the same. After proper normalization of \ref{wave_function_SC}, both expressions yield an identical wave function, and the apparent discrepancy vanishes.  

In obtaining the third line of \ref{wave_function_SC}, we have used the following identity: 
\begin{eqnarray}\label{squeezing_Displacement}
    \hat{S}_{\mathbf{k}}(r_{k},\phi_{k}) \hat{R}_{\mathbf{k}}(\theta_{k}) \hat{D}(\alpha_{\mathbf{k}}) \hat{D}(\alpha_{-\mathbf{k}}) &=& \hat{D}(\beta_{\mathbf{k}}) \hat{D}(\gamma_{\mathbf{k}}) \hat{S}_{\mathbf{k}}(r_{k},\phi_{k}) \hat{R}_{\mathbf{k}}(\theta_{k}) \hspace{0.06cm} . 
\end{eqnarray}

\section{Derivation of the time evolution operator}\label{appendix:C}

In this section, we derive the expression for the squeezing operator $\hat{S}_{\mathbf{k}}(r_{k}, \phi_{k})$ in terms of a product of three exponential operators, as given in~\ref{squeezing_fact}. To simplify the derivation, we consider a reduced form of the cosmological Hamiltonian given in~\ref{Hamiltonian_1}, associated with an un-partitioned Hilbert space:
\begin{eqnarray}\label{example_Hamiltonian}
    \hat{\mathcal{H}}_{k} &=& \frac{1}{2} \left[\left(\hat{a}^{2}_{\mathbf{k}} + \hat{a}^{\dagger2}_{\mathbf{k}} \right) + \frac{\lambda(t)}{2} \left(\hat{a}_{\mathbf{k}}\hat{a}^{\dagger}_{\mathbf{k}} + \hat{a}^{\dagger}_{\mathbf{k}}\hat{a}_{\mathbf{k}} \right) \right] \nonumber \\ 
    &=& \hat{A}_{\mathbf{k}} + \hat{B}_{\mathbf{k}} + \lambda(t) \hspace{0.07cm} \hat{C}_{\mathbf{k}} \hspace{0.1 cm} , 
\end{eqnarray}
where $\lambda(t)$ is an arbitrary time-dependent parameter of this quadratic Hamiltonian. The operators $\hat{A}_{\mathbf{k}}, \hat{B}_{\mathbf{k}}$, and $\hat{C}_{\mathbf{k}}$ are defined as follows:
\begin{eqnarray}
    && \hat{A}_{\mathbf{k}} \, = \, \frac{1}{2} \hspace{0.05cm} \hat{a}^{\dagger2}_{\mathbf{k}} \hspace{1.2cm} ; \hspace{0.4cm} \hat{B}_{\mathbf{k}} \, = \, \frac{1}{2} \hspace{0.05cm} \hat{a}^{2}_{\mathbf{k}} \hspace{1.0cm} ; \hspace{0.4cm} \hat{C}_{\mathbf{k}} \, = \, \frac{1}{4} \hspace{0.05cm} \left(\hat{a}_{\mathbf{k}}\hat{a}^{\dagger}_{\mathbf{k}} + \hat{a}^{\dagger}_{\mathbf{k}}\hat{a}_{\mathbf{k}} \right) \hspace{0.06cm} . 
\end{eqnarray}
These operators satisfy the following Lie algebra:
\begin{eqnarray}\label{Lie_algebra}
    [\hat{A}_{\mathbf{k}},\hat{B}_{\mathbf{k}}] = -2\hat{C}_{\mathbf{k}} \hspace{0.4cm} ; \hspace{0.4cm}  [\hat{C}_{\mathbf{k}},\hat{A}_{\mathbf{k}}] = \hspace{0.14cm} \hat{A}_{\mathbf{k}} \hspace{0.4cm} ; \hspace{0.4cm}  [\hat{C}_{\mathbf{k}},\hat{B}_{\mathbf{k}}] = -\hat{B}_{\mathbf{k}} \hspace{0.06cm} . 
\end{eqnarray}
The propagator of a quantum system is typically expressed as the exponential of its Hamiltonian. Since the Hamiltonian in ~\ref{example_Hamiltonian} is a linear combination of three operators obeying the above Lie algebra, we can utilize the Baker–Campbell–Hausdorff (BCH) formula to decompose the exponential of their sum.
Let us define a simplified propagator neglecting the $\hat{C}_{\mathbf{k}}$ term, as this does not contribute directly to the squeezing portion $\hat{S}_{\mathbf{k}}(r_{k}, \phi_{k})$ of the full time-evolution operator (see~\ref{time_evolution}):
\begin{eqnarray}\label{example_propagator_1}
    \hat{U}_{\mathbf{k}}(\zeta) &=& \exp[i\zeta \big(\hat{A}_{\mathbf{k}}+\hat{B}_{\mathbf{k}} \big)] \hspace{0.06cm} . 
\end{eqnarray}
According to the BCH decomposition, this can be factorized as:
\begin{equation}\label{example_propagator_2}
    \hat{U}_{\mathbf{k}}(\zeta) = \exp[ip(\zeta)\hat{A}_{\mathbf{k}}] \exp[iq(\zeta)\hat{C}_{\mathbf{k}}] \exp[ir(\zeta)\hat{B}_{\mathbf{k}}] \hspace{0.06cm} , 
\end{equation}
where $p(\zeta), q(\zeta)$ and $r(\zeta)$ are scalar functions to be determined.
To find them, we differentiate both sides with respect to $\zeta$. Starting with~\ref{example_propagator_1}:
\begin{eqnarray}\label{derivative_1}
    \frac{d\hat{U}_{\mathbf{k}}(\zeta)}{d\zeta} &=& i \big(\hat{A}_{\mathbf{k}}+\hat{B}_{\mathbf{k}} \big) \exp[i\zeta \big(\hat{A}_{\mathbf{k}}+\hat{B}_{\mathbf{k}} \big)] \hspace{0.06cm} . 
\end{eqnarray}
Differentiating~\ref{example_propagator_2} yield:
\begin{eqnarray}\label{example_intermed}
    \frac{d\hat{U}_{\mathbf{k}}(\zeta)}{d\zeta} &=& i \left[p'\hat{A}_{\mathbf{k}} + q'\bigg(e^{ip\hat{A}_{\mathbf{k}}}\hat{C}_{\mathbf{k}}e^{-ip\hat{A}_{\mathbf{k}}} \bigg) + r'\bigg(e^{ip\hat{A}_{\mathbf{k}}}e^{iq\hat{C}_{\mathbf{k}}}\hat{B}_{\mathbf{k}} \bigg)\bigg(e^{-iq\hat{C}_{\mathbf{k}}} e^{-ip\hat{A}_{\mathbf{k}}} e^{ip\hat{A}_{\mathbf{k}}} e^{iq\hat{C}_{\mathbf{k}}} \bigg) e^{ir\hat{B}_{\mathbf{k}}} \hat{U}^{-1}(\zeta) \right] \, \hat{U}_{\mathbf{k}}(\zeta)   \nonumber  \\ 
    &=& i \left[p'\hat{A}_{\mathbf{k}} + q'\bigg(e^{ip\hat{A}_{\mathbf{k}}}\hat{C}_{\mathbf{k}}e^{-ip\hat{A}_{\mathbf{k}}} \bigg) + r'\bigg(e^{ip\hat{A}_{\mathbf{k}}}e^{iq\hat{C}_{\mathbf{k}}}\hat{B}_{\mathbf{k}} e^{-iq\hat{C}_{\mathbf{k}}} e^{-ip\hat{A}_{\mathbf{k}}} \bigg) \right] \, \hat{U}_{\mathbf{k}}(\zeta) \hspace{0.06cm} .  
\end{eqnarray}
Using the Hadamard lemma:
\begin{eqnarray}\label{Hadamard}
    \exp\big(ip\hat{A}_{\mathbf{k}} \big) \hspace{0.1cm} \hat{C}_{\mathbf{k}} \hspace{0.1cm} \exp\big(-ip\hat{A}_{\mathbf{k}} \big) &=& \left(\hat{C}_{\mathbf{k}}-ip\hat{A}_{\mathbf{k}} \right) \hspace{0.06cm} , \\ 
    \exp\big(ip\hat{A}_{\mathbf{k}} \big) \hspace{0.07cm} \exp\big(iq\hat{C}_{\mathbf{k}} \big) \hspace{0.07cm}\hat{B}_{\mathbf{k}} \hspace{0.07cm} \exp\big(-iq\hat{C}_{\mathbf{k}} \big) \hspace{0.07cm} \exp\big(-ip\hat{A}_{\mathbf{k}} \big) &=& e^{-iq} \left(\hat{B}_{\mathbf{k}}-2ip \, \hat{C}_{\mathbf{k}}-p^{2}\hat{A}_{\mathbf{k}} \right) \hspace{0.06cm} . 
\end{eqnarray}
Substituting these into~\ref{example_intermed}, we get:
\begin{equation}\label{derivative_2}
    \frac{d\hat{U}_{\mathbf{k}}(\zeta)}{d\zeta} \, = \, i\left[\big(p'-ipq'-r'p^{2}e^{-iq} \big) \hat{A}_{\mathbf{k}} + \big(r'e^{-iq} \big)\hat{B}_{\mathbf{k}} + \big(q'-2ipr'e^{-iq} \big)\hat{C}_{\mathbf{k}} \right] \, \hat{U}_{\mathbf{k}}(\zeta) \hspace{0.06cm} . 
\end{equation}
Now, equating~\ref{derivative_1} and~\ref{derivative_2}, we obtain the following system of coupled differential equations:
\begin{eqnarray}
    \big(p'-ipq'-r'p^{2}e^{-iq} \big) &=& 1 \hspace{0.06cm} , \\ 
    r'e^{-iq} &=& 1 \hspace{0.06cm} , \\ 
    \big(q'-2i \hspace{0.1cm} pr'e^{-iq} \big) &=& 0 \hspace{0.06cm} . 
\end{eqnarray}
Solving this system of coupled differential equations yields the following solutions:
\begin{eqnarray}
    p(\zeta) &=& r(\zeta) = \hspace{0.21cm} \tanh{\zeta} \hspace{0.06cm} , \\ 
    q(\zeta) &=& 2i \hspace{0.05cm} \ln{(\cosh{\zeta})} \hspace{0.06cm} . 
\end{eqnarray}
Substituting back into~\ref{example_propagator_2}, we arrive at the final factorized form of the propagator:
\begin{eqnarray}
    \hat{U}_{\mathbf{k}}(\zeta) &=& \hspace{0.1cm} \exp[i\tanh{\zeta} \hspace{0.07cm} \hat{A}_{\mathbf{k}}] \hspace{0.1cm} \exp{-\ln{(\cosh^{2}{\zeta})} \hspace{0.07cm} [\hat{A}_{\mathbf{k}},\hat{B}_{\mathbf{k}}]} \hspace{0.1cm} \exp[i\tanh{\zeta} \hspace{0.07cm} \hat{B}_{\mathbf{k}}] \, . 
\end{eqnarray}

\section{Detailed computation of spin-xx correlation}\label{Spin-xx_calculation}

Here we provide the detailed computational steps required to evaluate the spin-xx correlation of the squeezed coherent state (for the wave function, see \ref{wave_function_SC}). The four integrals appearing in the expression for the spin-xx correlation (see \ref{Spin xx}) are labeled as follows:
\begin{eqnarray}\label{Spin-xx integral}
    I_{0} &=& \int^{\infty}_{0}\int^{\infty}_{0} dq_{\mathbf{k}}dq_{-\mathbf{k}} \hspace{0.1cm} \psi_{r}(q_{\mathbf{k}},q_{-\mathbf{k}}) \hspace{0.06cm} \psi^{*}_{r}(q_{\mathbf{k}},q_{-\mathbf{k}}) \hspace{0.1cm} , \\ 
    I_{1} &=& \int^{\infty}_{0} \int^{\infty}_{0} dq_{\mathbf{k}}dq_{-\mathbf{k}} \hspace{0.1cm} \psi_{r}(-q_{\mathbf{k}},-q_{-\mathbf{k}}) \hspace{0.06cm} \psi^{*}_{r}(-q_{\mathbf{k}},-q_{-\mathbf{k}}) \hspace{0.1cm} , \\ 
    I_{2} &=& \int^{\infty}_{0} \int^{\infty}_{0} dq_{\mathbf{k}} dq_{-\mathbf{k}} \hspace{0.1cm} \psi_{r}(q_{\mathbf{k}},-q_{-\mathbf{k}}) \hspace{0.06cm} \psi^{*}_{r}(q_{\mathbf{k}},-q_{-\mathbf{k}}) \hspace{0.1cm} , \\ 
    I_{3} &=& \int^{\infty}_{0} \int^{\infty}_{0} dq_{\mathbf{k}} dq_{-\mathbf{k}} \hspace{0.1cm} \psi_{r}(-q_{\mathbf{k}},q_{-\mathbf{k}}) \hspace{0.06cm} \psi^{*}_{r}(-q_{\mathbf{k}},q_{-\mathbf{k}}) \hspace{0.1cm} \label{spin-xx integral_3}. 
\end{eqnarray}

\begin{itemize}[\labelsep=0.2em ]

    \item \textbf{First Integral:} We begin with the evaluation of the first integral $I_0$, corresponding to the first term in the spin-xx correlation expression \ref{Spin xx}. To obtain an analytical result, we employ certain special functions, including the Error function (Erf), Owen’s T-function (OwnT), and the Signum function (Sign). The integral $I_0$ can be expressed as:  
\begin{eqnarray}\label{first_integral}
    I_{0} &=& \int^{\infty}_{0}\int^{\infty}_{0} dq_{\mathbf{k}}dq_{-\mathbf{k}} \hspace{0.1cm} \psi_{r}(q_{\mathbf{k}},q_{-\mathbf{k}}) \hspace{0.06cm} \psi^{*}_{r}(q_{\mathbf{k}},q_{-\mathbf{k}}) \nonumber \\
    &=& \abs{M_{0}}^{2} \int^{\infty}_{0} dw_{1} \hspace{0.1cm} e^{-g_{1}(w_{1}-x_{0})^{2}} \int^{w_{1}}_{-w_{1}} dw_{2} \hspace{0.1cm} e^{-g_{2}(w_{2}-y_{0})^{2}}  \nonumber \\ 
    &=& \frac{\abs{M_{0}}^{2}}{2} \sqrt{\frac{\pi}{g_{2}}} \int^{\infty}_{0} dw_{1} \hspace{0.1cm} e^{-g_{1}(w_{1}-x_{0})^{2}} \left\{\Erf{\left[\sqrt{g_{2}}(w_{1}-y_{0}) \right]} + \Erf{\left[\sqrt{g_{2}}(w_{1}+y_{0}) \right]} \right\} . 
\end{eqnarray}
Here, the prefactor $M_{0}$ is defined as
\begin{eqnarray}
    \abs{M_{0}}^{2} &=& \frac{\abs{N_{0}}^{2}}{2} \exp\left\{-2 \Re\left[A_{0}\big(C_{0}^{2} + D_{0}^{2} \big)\right] + g_{1} \, x_{0}^{2} + g_{2} \, y_{0}^{2} \, \right\} \hspace{0.06cm} . 
\end{eqnarray}
All the remaining coefficients such as $g_{1}, g_{2}, x_{0}, y_{0}$ and $N_{0}$, involved in the computation of first integral $I_{0}$, are already defined in the manuscript through the \ref{def-g_1}~-~\ref{def-M_0}. In \ref{first_integral}, we performed a change of variables from $(q_{\mathbf{k}},q_{-\mathbf{k}})$ to $(w_{1}, w_{2})$ , where 
\begin{equation}
    w_{1}:=(q_{\mathbf{k}}+q_{-\mathbf{k}})\,, \quad w_{2}:=(q_{\mathbf{k}}-q_{-\mathbf{k}}) \, . 
\end{equation}
 This transformation simplifies the domain of integration and allows us to isolate the dependence on each variable. After integrating over $w_2$, we are left with two nontrivial integrals involving the product of a Gaussian and the Error function, each with shifted arguments. The remaining one-dimensional integrals can be expressed in terms of Owen's T-function, Error function, and the Signum function, as follows~\cite{website}: 
\begin{eqnarray}
    \int^{\infty}_{0} dw_{1} \hspace{0.1cm} e^{-g_{1}(w_{1}-x_{0})^{2}} \Erf{\left[\sqrt{g_{2}} (w_{1}-y_{0}) \right]} \hspace{-0.2cm} &=& \hspace{-0.2cm} \frac{2\sqrt{\pi}}{\sqrt{g_{1}}} \left\{ \OwenT{\left[ - \sqrt{2 g_{1}}x_{0} \boldsymbol{;} \frac{\sqrt{g_{2}}y_{0}}{\sqrt{g_{1}}x_{0}} \right]} +  \frac{1}{4} \Erf{\left[ -\frac{\sqrt{g_{1}g_{2}}(y_{0} - x_{0})}{\sqrt{g_{1} + g_{2}}} \right]} + \right. \nonumber \\ 
    && \hspace{-0.5cm} \left. \OwenT{\left[ - \frac{\sqrt{2 g_{1}g_{2}}(y_{0}-x_{0})}{\sqrt{g_{1} + g_{2}}} \boldsymbol{;} \frac{\big(g_{1}x_{0}+g_{2}y_{0} \big)}{\sqrt{g_{1}g_{2}}(y_{0} - x_{0})} \right]} - \frac{1}{4} \Sign{\left[\sqrt{g_{2}} \left(\frac{y_{0}}{x_{0}} - 1 \right) \right]} \right\} , \\ 
    \int^{\infty}_{0} dw_{1} \hspace{0.1cm} e^{-g_{1}(w_{1}-x_{0})^{2}} \Erf{\left[\sqrt{g_{2}} (w_{1}+y_{0}) \right]} \hspace{-0.2cm} &=& \hspace{-0.2cm} \frac{2\sqrt{\pi}}{\sqrt{g_{1}}} \left\{\OwenT{\left[ - \sqrt{2 g_{1}}x_{0} \boldsymbol{;} -\frac{\sqrt{g_{2}}y_{0}}{\sqrt{g_{1}}x_{0}} \right]} + \frac{1}{4} \Erf{\left[\frac{\sqrt{g_{1}g_{2}}(y_{0} + x_{0})}{\sqrt{g_{1} + g_{2}}} \right]} + \right. \nonumber \\ 
    && \hspace{-0.5cm} \left. \OwenT{\left[ \frac{\sqrt{2 g_{1}g_{2}}(y_{0}+x_{0})}{\sqrt{g_{1} + g_{2}}} \boldsymbol{;} \frac{\big(g_{2}y_{0}-g_{1}x_{0} \big)}{\sqrt{g_{1}g_{2}}(y_{0} + x_{0})} \right]} - \frac{1}{4} \Sign{\left[ - \sqrt{g_{2}} \left(\frac{y_{0}}{x_{0}} + 1 \right) \right]} \right\} . \hspace{0.9cm}
\end{eqnarray}

    \item \textbf{Second Integral:} Next, we do the evaluation  second integral $I_{1}$. The primary difference between the first and second integral is lying in the arguments of the wave function. The arguments of the wave function $\psi_{r}(q_{\mathbf{k}},q_{-\mathbf{k}})$ present in the first integral, is replaced by another set of arguments $\psi_{r}(-q_{\mathbf{k}},-q_{-\mathbf{k}})$ in the second integral. The integration can be done as follows,   
\begin{eqnarray}\label{second_integral}
    I_{1} &=& \int^{\infty}_{0} \int^{\infty}_{0} dq_{\mathbf{k}}dq_{-\mathbf{k}} \hspace{0.1cm} \psi_{r}(-q_{\mathbf{k}},-q_{-\mathbf{k}}) \hspace{0.06cm} \psi^{*}_{r}(-q_{\mathbf{k}},-q_{-\mathbf{k}}) \nonumber \\ 
    &=& \abs{M_{0}}^{2} \int^{-\infty}_{0} dw_{1} \hspace{0.1cm} e^{-g_{1}(w_{1}-x_{0})^{2}} \int^{w_{1}}_{-w_{1}} dw_{2} \hspace{0.1cm} e^{-g_{2}(w_{2}-y_{0})^{2}} \nonumber \\ 
    &=& \frac{\abs{M_{0}}^{2}}{2} \sqrt{\frac{\pi}{g_{2}}} \int^{-\infty}_{0} dw_{1} \hspace{0.1cm} e^{-g_{1}(w_{1}-x_{0} )^{2}} \big[\Erf \{ {\sqrt{g_{2}}\big(w_{1}-y_{0} \big)} \} + \Erf \{ {\sqrt{g_{2}} \big(w_{1}+y_{0} \big)} \} \big] \hspace{0.06cm} . 
\end{eqnarray}
The only difference between the \ref{first_integral} and \ref{second_integral} lies in the upper limit of the $w_{1}$-integration which has been changed from $\infty$ to $-\infty$.  
Again following the same reference \cite{website}, the integrals \ref{second_integral} should be expressed in the following manner  
\begin{eqnarray}
    \int^{\infty}_{0} dw_{1} \hspace{0.1cm} e^{-g_{1}(w_{1}+x_{0})^{2}} \Erf{\{\sqrt{g_{2}} \left(w_{1}+y_{0} \right)\}} \hspace{-0.2cm} &=& \hspace{-0.2cm} \frac{2\sqrt{\pi}}{\sqrt{g_{1}}} \bigg[\OwenT{\bigg\{\hspace{-0.1cm} \sqrt{2 g_{1}}x_{0} \boldsymbol{;} \frac{\sqrt{g_{2}}y_{0}}{\sqrt{g_{1}}x_{0}} \bigg\}} + \frac{1}{4} \Erf{\bigg\{\hspace{-0.1cm} \frac{\sqrt{g_{1}g_{2}}(y_{0} - x_{0})}{\sqrt{g_{1} + g_{2}}} \bigg\}} + \nonumber \\ 
    && \hspace{-0.5cm} \OwenT{\bigg\{\hspace{-0.1cm} \frac{\sqrt{2 g_{1}g_{2}}(y_{0}-x_{0})}{\sqrt{g_{1} + g_{2}}} \boldsymbol{;} \frac{\big(g_{1}x_{0}+g_{2}y_{0} \big)}{\sqrt{g_{1}g_{2}}(y_{0} - x_{0})} \bigg\}} - \frac{1}{4} \Sign{\bigg\{\sqrt{g_{2}} \left(\frac{y_{0}}{x_{0}} - 1 \right) \bigg\}} \bigg] \hspace{0.06cm} , \nonumber \\ 
    \int^{\infty}_{0} dw_{1} \hspace{0.1cm} e^{-g_{1}(w_{1}+x_{0})^{2}} \Erf{\{\sqrt{g_{2}} \left(w_{1}-y_{0} \right)\}} \hspace{-0.2cm} &=& \hspace{-0.2cm} \frac{2\sqrt{\pi}}{\sqrt{g_{1}}} \bigg[\OwenT{\bigg\{\hspace{-0.1cm} \sqrt{2 g_{1}}x_{0} \boldsymbol{;} -\frac{\sqrt{g_{2}}y_{0}}{\sqrt{g_{1}}x_{0}} \bigg\}} + \frac{1}{4} \Erf{\bigg\{ \hspace{-0.1cm} - \frac{\sqrt{g_{1}g_{2}}(y_{0} + x_{0})}{\sqrt{g_{1} + g_{2}}} \bigg\}} + \nonumber \\ 
    && \hspace{-0.5cm} \OwenT{\bigg\{ \hspace{-0.1cm} - \hspace{-0.1cm} \frac{\sqrt{2 g_{1}g_{2}}(y_{0}+x_{0})}{\sqrt{g_{1} + g_{2}}} \boldsymbol{;} \frac{\big(g_{2}y_{0}-g_{1}x_{0} \big)}{\sqrt{g_{1}g_{2}}(y_{0} + x_{0})} \bigg\}} - \frac{1}{4} \Sign{\bigg\{ \hspace{-0.1cm} - \hspace{-0.1cm} \sqrt{g_{2}} \left(\frac{y_{0}}{x_{0}} + 1 \right) \bigg\}} \bigg] . \hspace{0.9cm}
\end{eqnarray}
    
    \item \textbf{Third Integral:} Our next job is to evaluate the third integral of spin-xx correlation \ref{Spin xx} where the second argument of the wave function $q_{-\mathbf{k}}$ is replaced by $-q_{-\mathbf{k}}$ without changing the first argument $q_{\mathbf{k}}$. The wave function, present in the third integral of spin-xx correlation, is $\psi_{r}(q_{\mathbf{k}}, -q_{-\mathbf{k}})$. Let us denote the integral by $I_{2}$ which can be done as follows,  
\begin{eqnarray}\label{third_integral}
    I_{2} &=& \int^{\infty}_{0} \int^{\infty}_{0} dq_{\mathbf{k}} dq_{-\mathbf{k}} \hspace{0.1cm} \psi_{r}(q_{\mathbf{k}},-q_{-\mathbf{k}}) \hspace{0.06cm} \psi^{*}_{r}(q_{\mathbf{k}},-q_{-\mathbf{k}}) \nonumber \\
    &=& -\abs{M_{0}}^{2} \int^{-\infty}_{0} dw_{2} \hspace{0.1cm} e^{-g_{2}(w_{2}-y_{0})^{2}} \hspace{0.1cm} \int^{w_{2}}_{-w_{2}} dw_{1} \hspace{0.1cm} e^{-g_{1}(w_{1}-x_{0})^{2}} \nonumber \\ 
    &=& -\frac{\abs{M_{0}}^{2}}{2} \sqrt{\frac{\pi}{g_{1}}} \int^{-\infty}_{0} dw_{2} \hspace{0.1cm} e^{-g_{2}(w_{2} - y_{0})^{2}} \hspace{0.1cm}  \big[\Erf{\{\sqrt{g_{1}}(w_{2}-x_{0}) \}} + \Erf{\{\sqrt{g_{1}}(w_{2}+x_{0}) \}} \big] \hspace{0.06cm} .
\end{eqnarray}
The primary difference between the first \ref{first_integral} and third \ref{third_integral} integral lies in the order of integration of the variables $w_{1}$ and $w_{2}$. Consequently, the limits of integration have been also changed in comparison with the first integral. Now, using the same reference \cite{website}, analytical results of the above integrals \ref{third_integral} can be expressed as 
\begin{eqnarray}
    \int^{\infty}_{0} dw_{2} \hspace{0.1cm} e^{-g_{2}(w_{2}+y_{0})^{2}} \Erf{\{\sqrt{g_{1}} \left(w_{2}+x_{0} \right) \}} \hspace{-0.2cm} &=& \hspace{-0.2cm} \frac{2\sqrt{\pi}}{\sqrt{g_{2}}} \bigg[\OwenT{\bigg\{\sqrt{2 g_{2}} y_{0} \boldsymbol{;} \frac{\sqrt{g_{1}}x_{0}}{\sqrt{g_{2}}y_{0}} \bigg\}} + \frac{1}{4} \Erf{\bigg\{\frac{\sqrt{g_{1}g_{2}}(x_{0}-y_{0})}{\sqrt{g_{1} + g_{2}}} \bigg\}} + \nonumber \\ 
    && \hspace{-0.5cm} \OwenT{\bigg\{ \frac{\sqrt{2 g_{1}g_{2}} (x_{0}-y_{0})}{\sqrt{g_{1} + g_{2}}} \boldsymbol{;} \frac{\left(g_{2}y_{0}+g_{1}x_{0} \right)}{\sqrt{g_{1}g_{2}} \left(x_{0}-y_{0} \right)} \bigg\}} - \frac{1}{4} \Sign{\bigg\{\sqrt{g_{1}} \left(\frac{x_{0}}{y_{0}} - 1 \right) \bigg\}} \bigg] \hspace{0.06cm} , \nonumber \\ 
    \int^{\infty}_{0} dw_{2} \hspace{0.1cm} e^{-g_{2}(w_{2}+y_{0})^{2}} \Erf{\{\sqrt{g_{1}} \left(w_{2}-x_{0} \right) \}} \hspace{-0.2cm} &=& \hspace{-0.2cm} \frac{2\sqrt{\pi}}{\sqrt{g_{2}}} \bigg[\OwenT{\bigg\{\sqrt{2 g_{2}} y_{0} \boldsymbol{;} -\frac{\sqrt{g_{1}}x_{0}}{\sqrt{g_{2}}y_{0}} \bigg\}} + \frac{1}{4} \Erf{\bigg\{\hspace{-0.1cm} - \frac{\sqrt{g_{1}g_{2}}(x_{0}+y_{0})}{\sqrt{g_{1} + g_{2}}} \bigg\}} + \nonumber \\ 
    && \hspace{-0.5cm} \OwenT{\bigg\{\hspace{-0.1cm} - \hspace{-0.1cm} \frac{\sqrt{2 g_{1}g_{2}} (x_{0}+y_{0})}{\sqrt{g_{1} + g_{2}}} \boldsymbol{;} \hspace{-0.06cm} - \frac{\left(g_{2}y_{0}-g_{1}x_{0} \right)}{\sqrt{g_{1}g_{2}} \left(x_{0}+y_{0} \right)} \bigg\}} - \frac{1}{4} \Sign{\bigg\{\hspace{-0.1cm} - \hspace{-0.1cm} \sqrt{g_{1}} \left(\frac{x_{0}}{y_{0}} + 1 \right) \bigg\}} \bigg] . \hspace{0.9cm}
\end{eqnarray}
    
    \item \textbf{Fourth Integral:} Our last task is to evaluate the fourth integral of spin-xx correlation \ref{Spin xx} where first argument of the wave function $q_{\mathbf{k}}$ is replaced by $-q_{\mathbf{k}}$ without changing the second argument $q_{-\mathbf{k}}$. The wave function, present in the fourth integral, is $\psi_{r}(-q_{\mathbf{k}}, q_{-\mathbf{k}})$. Let us denote it by $I_{3}$ which can be done as follows, 
\begin{eqnarray}\label{fourth_integral}
    I_{3} &=& \int^{\infty}_{0} \int^{\infty}_{0} dq_{\mathbf{k}} dq_{-\mathbf{k}} \hspace{0.1cm} \psi_{r}(-q_{\mathbf{k}},q_{-\mathbf{k}}) \psi^{*}_{r}(-q_{\mathbf{k}},q_{-\mathbf{k}}) \nonumber \\
    &=& -\abs{M_{0}}^{2} \int^{\infty}_{0} dw_{2} \hspace{0.1cm} e^{-g_{2}(w_{2}-y_{0})^{2}} \hspace{0.1cm} \int^{w_{2}}_{-w_{2}} dw_{1} \hspace{0.1cm} e^{-g_{1}(w_{1}-x_{0})^{2}} \nonumber \\ 
    &=& -\frac{\abs{M_{0}}^{2}}{2} \sqrt{\frac{\pi}{g_{1}}} \int^{\infty}_{0} dw_{2} \hspace{0.1cm} e^{-g_{2}(w_{2} - y_{0})^{2}} \big[\Erf{\{\sqrt{g_{1}}(w_{2}-x_{0}) \}} + \Erf{\{\sqrt{g_{1}}(w_{2}+x_{0}) \}} \big] \hspace{0.06cm} . 
\end{eqnarray}
The order of integration of the variables $w_{1}$ and $w_{2}$ is the only change that appears in the fourth integral \ref{fourth_integral} while comparing it with the first one \ref{first_integral}. 
According to this reference \cite{website}, analytical solutions of the above integrals are the following 
\begin{eqnarray}
    \int^{\infty}_{0} dw_{2} \hspace{0.1cm} e^{-g_{2}(w_{2}-y_{0})^{2}} \Erf{\{\sqrt{g_{1}} \left(w_{2}-x_{0} \right) \}} \hspace{-0.2cm} &=& \hspace{-0.2cm} \frac{2\sqrt{\pi}}{\sqrt{g_{2}}} \bigg[\OwenT{\bigg\{\hspace{-0.1cm} - \sqrt{2 g_{2}} y_{0} \boldsymbol{;} \frac{\sqrt{g_{1}}x_{0}}{\sqrt{g_{2}}y_{0}} \bigg\}} + \frac{1}{4} \Erf{\bigg\{ \hspace{-0.1cm} - \frac{\sqrt{g_{1}g_{2}}(x_{0}-y_{0})}{\sqrt{g_{1} + g_{2}}} \bigg\}} + \nonumber \\ 
    && \hspace{-0.5cm} \OwenT{\bigg\{\hspace{-0.1cm} - \frac{\sqrt{2 g_{1}g_{2}} (x_{0}-y_{0})}{\sqrt{g_{1} + g_{2}}} \boldsymbol{;} \frac{\left(g_{2}y_{0}+g_{1}x_{0} \right)}{\sqrt{g_{1}g_{2}} \left(x_{0}-y_{0} \right)} \bigg\}} - \frac{1}{4} \Sign{\bigg\{\sqrt{g_{1}} \left(\frac{x_{0}}{y_{0}} - 1 \right) \bigg\}} \bigg] \hspace{0.06cm} , \nonumber \\ 
    \int^{\infty}_{0} dw_{2} \hspace{0.1cm} e^{-g_{2}(w_{2}-y_{0})^{2}} \Erf{\{\sqrt{g_{1}} \left(w_{2}+x_{0} \right) \}} \hspace{-0.2cm} &=& \hspace{-0.2cm} \frac{2\sqrt{\pi}}{\sqrt{g_{2}}} \bigg[\OwenT{\bigg\{\hspace{-0.1cm} - \hspace{-0.1cm} \sqrt{2 g_{2}} y_{0} \boldsymbol{;} -\frac{\sqrt{g_{1}}x_{0}}{\sqrt{g_{2}}y_{0}} \bigg\}} + \frac{1}{4} \Erf{\bigg\{ \frac{\sqrt{g_{1}g_{2}}(x_{0}+y_{0})}{\sqrt{g_{1} + g_{2}}} \bigg\}} + \nonumber \\ 
    && \hspace{-0.5cm} \OwenT{\bigg\{\hspace{-0.1cm} \frac{\sqrt{2 g_{1}g_{2}} (x_{0}+y_{0})}{\sqrt{g_{1} + g_{2}}} \boldsymbol{;} \hspace{-0.06cm} - \frac{\left(g_{2}y_{0}-g_{1}x_{0} \right)}{\sqrt{g_{1}g_{2}} \left(x_{0}+y_{0} \right)} \bigg\}} - \frac{1}{4} \Sign{\bigg\{\hspace{-0.1cm} - \hspace{-0.1cm} \sqrt{g_{1}} \left(\frac{x_{0}}{y_{0}} + 1 \right) \bigg\}} \bigg] . \hspace{0.9cm}
\end{eqnarray}

\end{itemize}

Evaluation of rest of the integrals, denoted by $I_{1}, I_{2}$ and $I_{3}$, can be done in a similar fashion in terms of same set of special functions, Owen's T, Error and Signum function. Substituting the results of the remaining integrals, as given in the \ref{Spin-xx integral}~-~\ref{spin-xx integral_3}, in the spin-xx correlation \ref{Spin xx} of squeezed coherent state, one obtains the following simplified form  
\begin{eqnarray}\label{Spin-xx Correlation_2}
    \left\langle r,\alpha \right| \hat{S}_{x}(\mathbf{k})\hat{S}_{x}(-\mathbf{k}) \left| r,\alpha \right\rangle &=& \frac{4\pi\abs{M_{0}}^{2}}{\sqrt{g_{1} g_{2}}} \left\{\OwenT{\left[\hspace{0.1cm} \frac{\sqrt{2 g_{1} g_{2}} \, (y_{0}-x_{0})}{\sqrt{g_{1}+g_{2}}} \boldsymbol{;} \frac{\left(g_{1}x_{0}+g_{2}y_{0} \right)}{\sqrt{g_{1}g_{2}} \, (y_{0}-x_{0})} \right]} + \frac{1}{8} \Sign{\left[\sqrt{g_{2}} \left(\frac{y_{0}}{x_{0}}+1 \right) \right]} \right. \nonumber \\ 
    && \hspace{1.1cm} - \hspace{0.1cm} \OwenT{\left[\hspace{0.1cm} \frac{\sqrt{2 g_{1}g_{2}} \, (y_{0}+x_{0})}{\sqrt{g_{1}+g_{2}}} \boldsymbol{;} \frac{\left(g_{1}x_{0}-g_{2}y_{0} \right)}{\sqrt{g_{1}g_{2}} \, (y_{0}+x_{0})} \right]} - \frac{1}{8} \Sign{\left[\sqrt{g_{2}} \left(\frac{y_{0}}{x_{0}}-1 \right) \right]} \nonumber \\ 
    && \left. \hspace{1.2cm} + \hspace{0.1cm} \frac{1}{8} \Sign{\left[\sqrt{g_{1}} \left(\frac{x_{0}}{y_{0}}-1 \right) \right]} - \frac{1}{8} \Sign{\left[\sqrt{g_{1}} \left(\frac{x_{0}}{y_{0}}+1 \right) \right]} \right\} . 
\end{eqnarray}

\bibliographystyle{JHEP}
\bibliography{mybibliography-cosmology}

\end{document}